\documentclass[10pt,thmsa]{article}
\usepackage{graphics}
\input{tcilatex}
\begin{document}

\setcounter{page}{1}

\begin{center}
{\Large Functional Integral Approach in the Theory }

\vskip 0.2cm
%
{\Large of Color Superconductivity}{\footnote{%
Lectures at the VIth International School in Theoretical Physics, Vung Tau
27 December 1999-08 January 2000.}}

\vskip 0.4cm
%
\textbf{Nguyen Van Hieu}

\textit{Institute of Physics, NCST of Vietnam }

and

\textit{Faculty of Technology, Vietnam National University, Hanoi}

\vskip 1cm
%
\textbf{Abstract}
\end{center}

\begin{quotation}
In this series of lectures we present the functional integral method for
studying the superconducting pairing of quarks with the formation of the
diquarks as well as the quark-antiquark pairing in dense QCD. The dynamical
equations for the superconducting order parameters are the nonlinear
integral equations for the composite quantum fields describing the
quark-quark or quark-antiquark systems. These composite fields are the
bi-local fields if the pairing is generated by the gluon exchange while for
the instanton induced pairing interactions they are the local ones. The
expressions of the free energy densities are derived. The binding of three
quarks is also discussed.
\end{quotation}

\section{Introduction}

The superconducting pairing of quarks due to the gluon exchange in QCD with
the formation of the diquark condensate was proposed by Barrois$^{\left[
1\right] }$ and Frautschi$^{\left[ 2\right] }$ since more than two decades
and then studied by Bailin and Love$^{\left[ 3\right] }$, Donoglue and
Sateesh$^{\left[ 4\right] }$, Iwasaki and Iwado$^{\left[ 5\right] }.$
Recently, in a series of papers by Alford, Rajagopal and Wilczek$^{\left[
6\right] }$, Sch\"{a}fer and Wilczek$^{\left[ 7\right] }$, Rapp,
Sch\"{a}fer, Shuryak and Velkovsky$^{\left[ 8\right] }$, Evans, Hsu and
Schwetz$^{\left[ 9\right] }$, Son$^{\left[ 10\right] }$, Carter and Diakonov$%
^{\left[ 11\right] }$ and others$^{\left[ 12-21\right] }$ there arose a new
interest to the existence of the diquark Bose condensate in the QCD dense
matter-the color superconductivity. The connection between the color
superconductivity and the chiral phase transition in QCD was studied by
Berges and Rajagopal$^{\left[ 22\right] }$, Harada and Shibata$^{\left[
23\right] }$. There exists also the spontaneous parity violation, as it was
shown by Pisarski and Rischke$^{\left[ 24\right] }$.

For the study of many-body systems of relativistic particles with the
internal degrees of freedom as well as with the virtual creation and
annihilation of the particle-antiparticle pairs the functional integral
technique is a powerful mathematical tool. This method was applied to the
study of the color superconductivity as well as the quark-antiquark pairing$%
^{\left[ 25-28\right] }$.

In this series of lectures we present the basics of the functional integral
method for the study of the color superconductivity in QCD at finite
temperature and density (the so-called dense QCD or thermal QCD). As the
physical origin of the superconducting pairing of quarks we consider two
different commonly discussed mechanisms: the direct local four-fermion
interactions of the quark field and the quark-quark non-local interaction
due to the gluon exchange. The direct four-quark interaction might be
induced by the instantons$^{\left[ 11,12\right] }$.

We follow the general method of the functional integral approach in the
theory of superconductivity $^{\left[ 29,30\right] }$. We start from the
expressions of the partition functions of the systems of interacting quarks
and antiquarks with some local or non-local quark-quark interactions. Then
we introduce the local or bi-local composite fields describing the diquarks
or quark-antiquark pairs, establish the effective action of these composite
fields and derive their field equations. The order parameters of the ground
state of the diquark or quark-antiquark condensate are the expectation
values of the composite fields in the corresponding state of the system. Due
to the translational invariance of the ground state these expectation values
are coordinate-independent (for the local fields) or depend only on the
difference of the coordinates (for the bi-local ones). The equations for the
order parameters are the special cases of the field equations. The
expressions of the free energy density in the corresponding phases are also
derived.

We work in the imaginary time formalism and write briefly

\[
x=\left( \mathbf{x},\tau \right) ,\qquad \int dx=\int\limits_{0}^{\beta
}d\tau \int d\mathbf{x,}\qquad \beta =\frac{1}{kT}, 
\]
$k$ is the Boltzmann constant and $T$ is the temperature. We denote $\psi
_{A},$ $\overline{\psi }^{A}$ the quark field and its Dirac conjugate, where 
$A=\left( \alpha ,a,i\right) $ is the set consisting of the Dirac spinor
index $\alpha =1,2,3,4,$ the flavor index $i=1,2,3,....N_{f}$ and the color
one $a=1,2,3...N_{c}$ . The internal symmetry groups are assumed to be $%
SU\left( N_{f}\right) _{f}$ and $SU\left( N_{c}\right) _{c}$. The partition
function of the system of free quarks and antiquarks with the chemical
potential $\mu $ and at the temperature $T$ can be expressed in the form of
the functional integral

\begin{equation}
Z_{0}=\int \left[ D\psi \right] \left[ D\overline{\psi }\right] \exp \left\{
-\int dx\overline{\psi }^{A}\left( x\right) L_{A}^{B}\psi _{B}\left(
x\right) \right\}  \label{1}
\end{equation}
where

\begin{equation}
L_{A}^{B}=\delta _{a}^{b}\delta _{j}^{j}\left[ \gamma _{4}\left( \frac{%
\partial }{\partial \tau }-\mu \right) +\mathbf{\gamma \nabla }+M\right]
_{\alpha }^{\beta },  \label{2}
\end{equation}
and $M$ is the bare quark mass.

Introduce the generating functional

\begin{eqnarray}
Z_{0}\left[ \eta ,\overline{\eta }\right] &=&\int \left[ D\psi \right]
\left[ D\overline{\psi }\right] \exp \left\{ -\int dx\overline{\psi }%
^{A}\left( x\right) L_{A}^{B}\psi _{B}\left( x\right) \right\}  \nonumber \\
&&\exp \left\{ -\int dx\left[ \overline{\eta }^{A}\left( x\right) \psi
_{A}\left( x\right) +\overline{\psi }^{A}\left( x\right) \eta _{A}\left(
x\right) \right] \right\}  \label{3}
\end{eqnarray}
with anti-commuting parameters $\eta _{A}\left( x\right) $ and $\overline{%
\eta }^{A}\left( x\right) $. The $2n$-point Green functions are expressed in
terms of the functional derivatives of $Z_{0}\left[ \eta ,\overline{\eta }%
\right] $ at the special value $\eta _{A}\left( x\right) =\overline{\eta }%
^{A}\left( x\right) =0:$

\begin{eqnarray}
&&G_{A_{1},....A_{n}}^{B_{1},....B_{n}}\left(
x_{1},x_{2},...,x_{n};y_{1},y_{2},...,y_{n}\right) 
\begin{array}{l}
=
\end{array}
\text{ \qquad }  \nonumber \\
&=&\left\langle T\left\{ \psi _{A_{1}}\left( x_{1}\right) \psi
_{A_{2}}\left( x_{2}\right) ...\psi _{A_{n}}\left( x_{n}\right) \overline{%
\psi }^{B_{n}}\left( y_{n}\right) ...\overline{\psi }^{B_{2}}\left(
y_{2}\right) \overline{\psi }^{B_{1}}\left( y_{1}\right) \right\}
\right\rangle  \nonumber \\
&=&\left( -1\right) ^{n}\frac{1}{Z_{0}}\frac{\delta ^{2n}Z_{0}\left[ \eta ,%
\overline{\eta }\right] }{\delta \overline{\eta }^{A_{1}}\left( x_{1}\right)
...\delta \overline{\eta }^{A_{n}}\left( x_{n}\right) \delta \eta
_{B_{n}}\left( y_{n}\right) ...\delta \eta _{B_{1}}\left( y_{1}\right) }\mid
_{\eta =\overline{\eta }=0}.  \label{4}
\end{eqnarray}
In particular

\begin{eqnarray}
G_{A}^{B}\left( x;y\right) &=&G_{A}^{B}\left( x-y\right) =\left\langle
T\left\{ \psi _{A}\left( x\right) \overline{\psi }^{B}\left( y\right)
\right\} \right\rangle  \nonumber \\
&=&-\frac{1}{Z_{0}}\frac{\delta ^{2}Z_{0}\left[ \eta ,\overline{\eta }%
\right] }{\delta \overline{\eta }^{A}\left( x\right) \delta \eta _{B}\left(
y\right) }\mid _{\eta =\overline{\eta }=0}.  \label{5}
\end{eqnarray}

Denote

\begin{equation}
S_{A}^{B}\left( x-y\right) =\delta _{a}^{b}\delta _{i}^{j}S_{\alpha }^{\beta
}\left( x-y\right) =\delta _{a}^{b}\delta _{i}^{j}S_{\alpha }^{\beta }\left( 
\mathbf{x}-\mathbf{y,}\tau -\sigma \right) ,  \label{6}
\end{equation}
\[
S_{\alpha }^{\beta }\left( \mathbf{x}-\mathbf{y,}\tau -\sigma \right)
=S_{\alpha \beta }\left( \mathbf{x}-\mathbf{y,}\tau -\sigma \right) 
\]
the solution of the equation

\begin{equation}
L_{A}^{B}S_{B}^{C}\left( x-y\right) =\delta _{A}^{C}\delta \left( x-y\right)
,  \label{7}
\end{equation}

\[
x=\left( \mathbf{x},\tau \right) ,\quad y=\left( \mathbf{y},\sigma \right)
,\quad \delta \left( x-y\right) =\delta \left( \mathbf{x}-\mathbf{y}\right)
\delta \left( \tau -\sigma \right) , 
\]

\begin{equation}
\left[ \gamma _{4}\left( \frac{\partial }{\partial \tau }-\mu \right) +%
\mathbf{\gamma \nabla }+M\right] _{\alpha \beta }S_{\beta \gamma }\left( 
\mathbf{x}-\mathbf{y,}\tau -\sigma \right) =\delta _{\alpha \gamma }\delta
\left( \mathbf{x}-\mathbf{y}\right) \delta \left( \tau -\sigma \right) .
\label{8}
\end{equation}
Shifting the functional integration variables

\begin{eqnarray*}
\psi _{B}\left( x\right) &\rightarrow &\psi _{B}\left( x\right) +\int
dyS_{B}^{D}\left( x-y\right) \eta _{D}\left( y\right) , \\
\overline{\psi }^{A}\left( x\right) &\rightarrow &\overline{\psi }^{A}\left(
x\right) +\int dz\overline{\eta }^{C}\left( z\right) S_{C}^{A}\left(
z-x\right)
\end{eqnarray*}
in the r.h.s. of the formula (1), we derive the explicit expression of the
generating functional (3)

\begin{equation}
Z_{0}\left[ \eta ,\overline{\eta }\right] =Z_{0}\exp \left\{ \int dx\int dy%
\overline{\eta }^{A}\left( x\right) S_{A}^{B}\left( x-y\right) \eta
_{B}\left( y\right) \right\} .  \label{9}
\end{equation}
Substituting this expression into the r.h.s. of the formula (5), we obtain
the two-point Green function

\begin{equation}
G_{A}^{B}\left( x-y\right) =\left\langle T\left\{ \psi _{A}\left( x\right) 
\overline{\psi }^{B}\left( y\right) \right\} \right\rangle =S_{A}^{B}\left(
x-y\right) .  \label{10}
\end{equation}
Similarly, from the formulae (4) and (9) it follows the Wick theorem for the 
$2n$-point Green functions of the free fermionic fields.

Introduce the Fourier transform $\widetilde{S}_{\alpha \beta }\left(
p\right) =\widetilde{S}_{\alpha \beta }\left( \mathbf{p},\varepsilon
_{m}\right) $ of $S_{\alpha \beta }\left( x\right) =S_{\alpha \beta }\left( 
\mathbf{x},\tau \right) :$

\begin{eqnarray}
S_{\alpha \beta }\left( \mathbf{x},\tau \right) &=&\frac{1}{\beta }%
\sum\limits_{m}e^{i\varepsilon _{m}t}\frac{1}{\left( 2\pi \right) ^{3}}\int d%
\mathbf{p\,\,}e^{i\mathbf{px}}\,\,\widetilde{S}_{\alpha \beta }\left( 
\mathbf{p},\varepsilon _{m}\right)  \label{11} \\
\varepsilon _{m} &=&\left( 2m+1\right) \frac{\pi }{2},  \nonumber
\end{eqnarray}
$m$ being the integers $m=0,\pm 1,\pm 2...$ From the equation (8) it follows
that

\begin{equation}
\widetilde{S}_{\alpha \beta }\left( p\right) =\left( \frac{1}{i\widehat{p}+M}%
\right) _{\alpha \beta }=\frac{\left( -i\widehat{p}+M\right) _{\alpha \beta }%
}{p^{2}+M}  \label{12}
\end{equation}
with the notations

\begin{equation}
\widehat{p}=\left( \varepsilon _{m}+i\mu \right) \gamma _{4}+\mathbf{\gamma
p,\qquad }p^{2}=\left( \varepsilon _{m}+i\mu \right) ^{2}+\mathbf{p}^{2}.
\label{13}
\end{equation}
In the calculations we shall use also the expression of $\widetilde{S}%
_{\alpha \beta }\left( -p\right) =\widetilde{S}_{\alpha \beta }\left( -%
\mathbf{p,-}\varepsilon _{m}\right) .$ For the convenience we write it in
the form

\begin{equation}
\widetilde{S}_{\alpha \beta }\left( -p\right) =\left( \frac{1}{-i\widehat{p}%
^{\prime }+M}\right) _{\alpha \beta }=\frac{\left( i\widehat{p}^{\prime
}+M\right) _{\alpha \beta }}{p^{2}+M},  \label{14}
\end{equation}
where

\begin{equation}
\widehat{p}^{\prime }=\left( \varepsilon _{m}-i\mu \right) \gamma _{4}+%
\mathbf{\gamma p,\qquad }p^{\prime \,\,2}=\left( \varepsilon _{m}-i\mu
\right) ^{2}+\mathbf{p}^{2}.  \label{15}
\end{equation}

To study the color superconductivity we consider the quark-quark pairing in
QCD with the formation of the diquark condensate. Then we investigate the
quark-antiquark pairing. The corresponding composite fields are the meson
ones. We discuss also the possibility to extend our reasonings to the study
of the binding of three quarks. The composite particles in this case are the
baryons.

\section{Quark-Quark Pairing}

For the simplicity in writing the formulae we begin our study by considering
the superconducting quark-quark pairing due to some direct four-fermion
coupling of quarks with the interaction Lagrangian

\begin{eqnarray}
L_{int} &=&\frac{1}{2}\overline{\psi }^{A}(x)\overline{\psi }%
^{C}(x)V_{CA}^{BD}\psi _{D}(x)\psi _{B}(x),  \nonumber \\
V_{AC}^{BD}\hspace{0in} &=&V_{CA}^{DB}=-V_{CA}^{BD}=-V_{AC}^{DB}.  \label{16}
\end{eqnarray}
The partition function of the system equals

\begin{eqnarray}
Z &=&\int [D\psi ][D\overline{\psi }]\exp \left\{ -\int dx\overline{\psi }%
^{A}(x)L_{A}^{B}\psi _{B}(x)\right\}  \nonumber \\
&&.\exp \left\{ \frac{1}{2}\int dx\overline{\psi }^{A}(x)\overline{\psi }%
^{C}(x)V_{CA}^{BD}\hspace{0in}\psi _{D}(x)\psi _{B}(x)\right\} .  \label{17}
\end{eqnarray}
Introduce the antisymmetric bi-spinor local fields $\Phi _{CA}\left(
x\right) ,\,\overline{\Phi }^{AC}\left( x\right) $

\begin{equation}
\overline{\Phi }^{CA}\left( x\right) =-\overline{\Phi }^{AC}\left( x\right)
,\Phi _{AC}\left( x\right) =-\Phi _{CA}\left( x\right) ,\,\,\,\,\,
\label{18}
\end{equation}
and the functional integral

\begin{equation}
Z_{0}^{\Phi }=\int \left[ D\Phi \right] \left[ D\overline{\Phi }\right] \exp
\left\{ -\frac{1}{2}\int dx\overline{\Phi }^{AC}\left( x\right)
V_{CA}^{BD}\left( x-y\right) \Phi _{DB}\left( x\right) \right\} .  \label{19}
\end{equation}
By shifting the functional integration variables

\begin{eqnarray*}
\Phi _{DB}\left( x\right) &\rightarrow &\Phi _{DB}\left( x\right) +\psi
_{D}\left( x\right) \psi _{B}\left( x\right) , \\
\overline{\Phi }^{AC}\left( x\right) &\rightarrow &\overline{\Phi }%
^{AC}\left( x\right) +\overline{\psi }^{A}\left( x\right) \overline{\psi }%
^{C}\left( x\right) ,
\end{eqnarray*}
we establish the Hubbard-Stratonovich transformation

\begin{eqnarray}
&&\exp \left\{ \frac{1}{2}\int dx\overline{\psi }^{A}\left( x\right) 
\overline{\psi }^{C}\left( x\right) V_{CA}^{BD}\psi _{D}\left( x\right) \psi
_{B}\left( x\right) \right\}  \label{20} \\
\qquad &=&\frac{1}{Z_{0}^{\Phi }}\int \left[ D\Phi \right] \left[ D\overline{%
\Phi }\right] \exp \left\{ -\frac{1}{2}\int dx\overline{\Phi }^{AC}\left(
x\right) V_{CA}^{BD}\Phi _{DB}\left( x\right) \right\} {}  \nonumber \\
&&\exp \left\{ -\frac{1}{2}\int dx\left[ \overline{\Delta }^{BD}\left(
x\right) \psi _{D}\left( x\right) \psi _{B}\left( x\right) +\overline{\psi }%
^{A}\left( x\right) \overline{\psi }^{C}\left( x\right) \Delta _{CA}\left(
x\right) \right] \right\} , \hskip 0.6cm \nonumber
\end{eqnarray}
where 
\begin{equation}
\overline{\Delta }^{BD}\left( x\right) =\overline{\Phi }^{AC}\left( x\right)
V_{CA}^{BD},\qquad \Delta _{CA}\left( x\right) =V_{CA}^{BD}\Phi _{DB}\left(
x\right) ,  \label{21}
\end{equation}
\begin{equation}
\overline{\Delta }^{DB}\left( x\right) =-\overline{\Delta }^{BD}\left(
x\right) ,\qquad \Delta _{AC}\left( x\right) =-\Delta _{CA}\left( x\right) ,
\label{22}
\end{equation}
and rewrite the partition function (17) in the form

\begin{equation}
Z=\frac{1}{Z_{0}^{\Phi }}\int \left[ D\Phi \right] \left[ D\overline{\Phi }%
\right] \exp \left\{ S_{\text{eff}}\left[ \Phi ,\overline{\Phi }\right]
\right\}  \label{23}
\end{equation}
with the effective action

\begin{equation}
S_{\text{eff}}\left[ \Phi ,\overline{\Phi }\right] =-\frac{1}{2}\int dx%
\overline{\Phi }^{BD}\left( x\right) V_{DB}^{AC}\Phi _{CA}\left( x\right)
+W\left[ \Delta ,\overline{\Delta }\right] ,  \label{24}
\end{equation}
\begin{eqnarray}
&&\exp \left\{ W\left[ \Delta ,\overline{\Delta }\right] \right\} 
\begin{array}{l}
=
\end{array}
\nonumber \\
&=&1+\sum_{n=1}^{\infty }\Gamma ^{\left( 2n\right) }\left[ \Delta ,\overline{%
\Delta }\right]  \nonumber \\
&=&\left\langle \text{T}\left[ \exp \left\{ -\frac{1}{2}\int dx\left[ 
\overline{\psi }^{A}\left( x\right) \overline{\psi }^{C}\left( x\right)
\Delta _{CA}\left( x\right) +\overline{\Delta }^{BD}\left( x\right) \psi
_{D}\left( x\right) \psi _{B}\left( x\right) \right] \right\} \right]
\right\rangle .\qquad
\end{eqnarray}
Calculations give 
\begin{eqnarray}
\Gamma ^{\left( 2\right) }\left[ \Delta ,\overline{\Delta }\right]
&=&W^{\left( 2\right) }\left[ \Delta ,\overline{\Delta }\right] =  \label{26}
\\
&=&\frac{1}{2}\int dx_{1}\int dx_{2}\overline{\Delta }^{A_{1}C_{1}}\left(
x_{1}\right) S_{C_{1}}^{C_{2}}\left( x_{1}-x_{2}\right) \Delta
_{C_{2}A_{2}}\left( x_{2}\right) S_{A_{1}}^{\text{T}A_{2}}\left(
x_{2}-x_{1}\right) ,\qquad  \nonumber  \label{26}
\end{eqnarray}
\begin{equation}
\Gamma ^{\left( 4\right) }\left[ \Delta ,\overline{\Delta }\right] =\frac{1}{%
2!}\left( W^{\left( 2\right) }\left[ \Delta ,\overline{\Delta }\right]
\right) ^{2}+W^{\left( 4\right) }\left[ \Delta ,\overline{\Delta }\right]
,\qquad \qquad  \label{27}
\end{equation}
\begin{eqnarray}
W^{\left( 4\right) }\left[ \Delta ,\overline{\Delta }\right] &=&-\frac{1}{4}%
\int dx_{1}...\,\int dx_{4}\overline{\Delta }^{A_{1}C_{1}}\left(
x_{1}\right) S_{C_{1}}^{C_{2}}\left( x_{1}-x_{2}\right) \Delta
_{C_{2}A_{2}}\left( x_{2}\right) S_{A_{3}}^{\text{T}A_{2}}\left(
x_{2}-x_{3}\right)  \nonumber \\
&&\overline{\Delta }^{A_{3}C_{3}}\left( x_{3}\right) S_{C_{3}}^{C_{4}}\left(
x_{3}-x_{4}\right) \Delta _{C_{4}A_{4}}\left( x_{4}\right) S_{A_{1}}^{\text{T%
}A_{4}}\left( x_{4}-x_{1}\right) ,\qquad  \label{28}
\end{eqnarray}
\begin{equation}
\Gamma ^{\left( 6\right) }\left[ \Delta ,\overline{\Delta }\right] =\frac{1}{%
3!}\left( W^{\left( 2\right) }\left[ \Delta ,\overline{\Delta }\right]
\right) ^{3}+W^{\left( 2\right) }\left[ \Delta ,\overline{\Delta }\right]
W^{\left( 4\right) }\left[ \Delta ,\overline{\Delta }\right] +W^{\left(
6\right) }\left[ \Delta ,\overline{\Delta }\right] ,  \label{29}
\end{equation}
\begin{eqnarray}
W^{\left( 6\right) }\left[ \Delta ,\overline{\Delta }\right] &=&\frac{1}{6}%
\int dx_{1}...\,\int dx_{6}\overline{\Delta }^{A_{1}C_{1}}\left(
x_{1}\right) S_{C_{1}}^{C_{2}}\left( x_{1}-x_{2}\right) \Delta
_{C_{2}A_{2}}\left( x_{2}\right) S_{A_{3}}^{\text{T}A_{2}}\left(
x_{2}-x_{3}\right)  \nonumber \\
&&...\,\,\overline{\Delta }^{A_{5}C_{5}}\left( x_{5}\right)
S_{C_{3}}^{C_{4}}\left( x_{5}-x_{6}\right) \Delta _{C_{6}A_{6}}\left(
x_{6}\right) S_{A_{1}}^{\text{T}A_{6}}\left( x_{6}-x_{1}\right) ,  \label{30}
\\
................... &&  \nonumber
\end{eqnarray}
with 
\[
S_{B}^{\text{T}A}\left( x_{1}-x_{2}\right) =S_{B}^{A}\left(
x_{2}-x_{1}\right) . 
\]
We have then 
\begin{equation}
W\left[ \Delta ,\overline{\Delta }\right] =\sum_{n=1}^{\infty }W^{\left(
2n\right) }\left[ \Delta ,\overline{\Delta }\right] .  \label{31}
\end{equation}

From the variational principle for the effective action we derive the field
equation 
\begin{equation}
\frac{1}{2}\Delta _{CA}\left( x\right) =V_{CA}^{BD}\sum_{n=1}^{\infty }\frac{%
\delta W^{\left( 2n\right) }\left[ \Delta ,\overline{\Delta }\right] }{%
\delta \overline{\Delta }^{BD}\left( x\right) }.  \label{32}
\end{equation}
Using the explicit expressions of $W^{\left( 2n\right) }\left[ \Delta ,%
\overline{\Delta }\right] $, we obtain 
\begin{eqnarray}
\Delta _{CA}\left( x\right) &=&V_{CA}^{BD}\left\{ \int
dx_{2}S_{D}^{C_{2}}\left( x-x_{2}\right) \Delta _{C_{2}A_{2}}\left(
x_{2}\right) S_{B}^{\text{T}A_{2}}\left( x_{2}-x\right) \right.  \nonumber \\
&&-\int dx_{2}...\,\int dx_{4}S_{D}^{C_{2}}\left( x-x_{2}\right) \Delta
_{C_{2}A_{2}}\left( x_{2}\right) S_{A_{3}}^{\text{T}A_{2}}\left(
x_{2}-x_{3}\right)  \nonumber \\
&&\qquad \overline{\Delta }^{A_{3}C_{3}}\left( x_{3}\right)
S_{C_{3}}^{C_{4}}\left( x_{3}-x_{4}\right) \Delta _{C_{4}A_{4}}\left(
x_{4}\right) S_{B}^{\text{T}A_{4}}\left( x_{4}-x\right)  \nonumber \\
&&+\int dx_{2}...\,\int dx_{6}S_{D}^{C_{2}}\left( x-x_{2}\right) \Delta
_{C_{2}A_{2}}\left( x_{2}\right) S_{A_{3}}^{\text{T}A_{2}}\left(
x_{2}-x_{3}\right)  \label{33} \\
&&\qquad \overline{\Delta }^{A_{3}C_{3}}\left( x_{3}\right)
S_{C_{3}}^{C_{4}}\left( x_{3}-x_{4}\right) ...\,\Delta _{C_{6}A_{6}}\left(
x_{6}\right) S_{B}^{\text{T}A_{6}}\left( x_{6}-x\right)  \nonumber \\
&&-\left. ...\hspace{3in}\QDATOP{\,}{\,}\right\} .  \nonumber
\end{eqnarray}
In the special class of the constant solutions 
\begin{equation}
\Delta _{CA}\left( x\right) =\Delta _{CA}=V_{CA}^{BD}\Phi _{DB},\qquad 
\overline{\Delta }^{BD}\left( x\right) =\overline{\Delta }^{BD}=\overline{%
\Phi }^{AC}V_{CA}^{BD},  \label{34}
\end{equation}
we have the extended BCS gap equation 
\begin{eqnarray}
\Delta _{CA} &=&V_{CA}^{BD}\frac{1}{\beta }\sum_{m}\frac{1}{\left( 2\pi
\right) ^{3}}\int d\mathbf{p}  \nonumber \\
&&\left[ \widetilde{S}\left( \mathbf{p},\varepsilon _{m}\right) \Delta \,%
\widetilde{S}^{\text{T}}\left( -\mathbf{p},-\varepsilon _{m}\right) \frac{1}{%
1+\overline{\Delta }\,\widetilde{S}\left( \mathbf{p},\varepsilon _{m}\right)
\Delta \,\widetilde{S}^{\text{T}}\left( -\mathbf{p},-\varepsilon _{m}\right) 
}\right] _{DB},  \hskip 0.6cm \label{35}
\end{eqnarray}
where $\widetilde{S}\left( \mathbf{p},\varepsilon _{m}\right) ,$ $\,%
\widetilde{S}^{\text{T}}\left( -\mathbf{p},-\varepsilon _{m}\right) ,$ $%
\Delta $ and $\overline{\Delta }$ are the matrices with the elements $%
\widetilde{S}_{A}^{B}\left( \mathbf{p},\varepsilon _{m}\right) ,\linebreak \,$ $%
\widetilde{S}_{A}^{\text{T}B}\left( -\mathbf{p},-\varepsilon _{m}\right)
,\,\Delta _{CA},$ and $\,\overline{\Delta }^{AC}$

At the values of the superconducting order parameters $\Delta _{CA}$ and $%
\overline{\Delta }^{BD}$ satisfying the extended BCS gap equation the
effective action equals 
\begin{eqnarray}
S_{\text{eff}}\left[ \Phi ,\overline{\Phi }\right] &=&\left( \frac{1}{2}-%
\frac{1}{4}\right) \int dx_{1}...\,\int dx_{4}\limfunc{Tr}\left[ \overline{%
\Delta }S\left( x_{1}-x_{2}\right) \Delta S^{\text{T}}\left(
x_{2}-x_{3}\right) \right.  \nonumber \\
&&\qquad \left. \overline{\Delta }\,\,S\left( x_{3}-x_{4}\right) \Delta \,S^{%
\text{T}}\left( x_{4}-x_{1}\right) \right]  \label{36} \\
&&-\left( \frac{1}{2}-\frac{1}{6}\right) \int dx_{1}...\,\int dx_{6}\limfunc{%
Tr}\left[ \overline{\Delta }\,S\left( x_{1}-x_{2}\right) \Delta \,S^{\text{T}%
}\left( x_{2}-x_{3}\right) \,\right.  \nonumber  \\
&&\qquad \left. ...\,\overline{\Delta }\,S\left( x_{5}-x_{6}\right) \Delta
\,S^{\text{T}}\left( x_{6}-x_{1}\right) \right]  \nonumber \\
&&+...  \nonumber
\end{eqnarray}
Denote $F\left[ \mathbf{x};\Delta \right] $ the free energy density of the
condensate. The effective action is expressed in terms of this free energy
density in the following manner

\begin{equation}
S_{\text{eff}}\left[ \Phi ,\overline{\Phi }\right] =-\beta \int d\mathbf{x}%
F\left[ \mathbf{x};\Delta \right] .  \label{37}
\end{equation}
Comparing (36) with (37), we obtain

\begin{eqnarray}
F\left[ \mathbf{x};\,\Delta \right] &=&F\left[ \Delta \right] =-\frac{1}{%
\beta }\sum_{m}\frac{1}{\left( 2\pi \right) ^{3}}\int d\mathbf{p}  \nonumber
\\
&&\limfunc{Tr}\left\{ \left( \frac{1}{2}-\frac{1}{4}\right) \left[ \overline{%
\Delta }\,\widetilde{S}\left( \mathbf{p},\varepsilon _{m}\right) \Delta \,%
\widetilde{S}^{\text{T}}\left( -\mathbf{p},-\varepsilon _{m}\right) \right]
^{2}\right.  \nonumber \\
&&-\left( \frac{1}{2}-\frac{1}{6}\right) \left[ \overline{\Delta }\,%
\widetilde{S}\left( \mathbf{p},\varepsilon _{m}\right) \Delta \,\widetilde{S}%
^{\text{T}}\left( -\mathbf{p},-\varepsilon _{m}\right) \right] ^{3}
\label{38} \\
&&+\left. \left( \frac{1}{2}-\frac{1}{8}\right) \left[ \overline{\Delta }\,%
\widetilde{S}\left( \mathbf{p},\varepsilon _{m}\right) \Delta \,\widetilde{S}%
^{\text{T}}\left( -\mathbf{p},-\varepsilon _{m}\right) \right]
^{4}\,-...\right\} .  \nonumber
\end{eqnarray}
Summing up the infinite series, we write the r.h.s of the formula (38) in
the compact form 
\begin{eqnarray}
F\left[ \mathbf{x};\,\Delta \right] &=&\frac{1}{\beta }\sum\limits_{m}\frac{1%
}{\left( 2\pi \right) ^{3}}\int d\mathbf{p}\frac{1}{2}\limfunc{Tr}\left[ 
\widetilde{S}\left( \mathbf{p},\varepsilon _{m}\right) \Delta \widetilde{S}%
^{T}\left( -\mathbf{p},-\varepsilon _{m}\right) \overline{\Delta }\right. 
\nonumber \\
&&\left. \left\{ \frac{1}{1+\widetilde{S}\left( \mathbf{p},\varepsilon
_{m}\right) \Delta \widetilde{S}^{T}\left( -\mathbf{p},-\varepsilon
_{m}\right) \overline{\Delta }}\right. \right.  \label{39} \\
&&\left. \left. -\int\limits_{0}^{1}d\alpha \frac{1}{1+\alpha \widetilde{S}%
\left( \mathbf{p},\varepsilon _{m}\right) \Delta \widetilde{S}^{T}\left( -%
\mathbf{p},-\varepsilon _{m}\right) \overline{\Delta }}\right\} \right] . 
\nonumber
\end{eqnarray}

Let us discuss the general form of the superconducting order parameters.
Consider first the constants $\Delta _{AC}$. We have 
\begin{equation}
\Delta _{AC}=\Delta _{\left( ai\alpha \right) \left( ck\gamma \right)
}=\left( \gamma _{5}C\right) _{\alpha \gamma }\Delta _{\left( ai\right)
\left( ck\right) }^{S}+\left( C\right) _{\alpha \gamma }\Delta _{\left(
ai\right) \left( ck\right) }^{P},  \label{40}
\end{equation}
where $\Delta _{\left( ai\right) \left( ck\right) }^{S}$ are the scalar
constants while the $\Delta _{\left( ai\right) \left( ck\right) }^{P}$ are
the pseudoscalar ones. If the parity is conserved, then all pseudoscalar
constants $\Delta _{\left( ai\right) \left( ck\right) }^{P}$ must be zero.
The existence of non-vanishing pseudoscalar constants $\Delta _{\left(
ai\right) \left( ck\right) }^{P}$ would signify the spontaneous breaking of
the parity conservation. Because of the condition (22) the constants $\Delta
_{\left( ai\right) \left( ck\right) }^{S}$ and $\Delta _{\left( ai\right)
\left( ck\right) }^{P}$ must have the property 
\begin{equation}
\Delta _{\left( ck\right) \left( ai\right) }^{S}=\Delta _{\left( ai\right)
\left( ck\right) }^{S},\qquad \Delta _{\left( ck\right) \left( ai\right)
}^{P}=\Delta _{\left( ai\right) \left( ck\right) }^{P}.  \label{41}
\end{equation}
This means, in particular, that if they are symmetric (antisymmetric) with
respect to the flavor indices $i$ and $j$, they must be also symmetric
(antisymmetric) with respect to the color ones $a$ and $b$.

For the study of the quark-quark pairing due to the gluon exchange we start
from the partition function in the form

\begin{eqnarray}
Z &=&\int [D\psi ][D\overline{\psi }]\exp \left\{ -\int dx\overline{\psi }%
^{A}(x)L_{A}^{B}\psi _{B}(x)\right\}  \nonumber \\
&&.\exp \left\{ \frac{1}{2}\int dx\int dy\overline{\psi }^{A}(x)\overline{%
\psi }^{C}(y)V_{CA}^{BD}\hspace{0in}\left( x-y\right) \psi _{D}(y)\psi
_{B}(x)\right\} ,  \label{42}
\end{eqnarray}
where

\begin{eqnarray}
V_{CA}^{BD}\hspace{0in}\left( x-y\right) &=&-\frac{g^{2}}{2\pi ^{2}}
\sum\limits_{I}\left( \gamma _{\mu }\otimes \lambda _{I}\right)
_{A}^{B}\left( \gamma _{\mu }\otimes \lambda _{I}\right) _{C}^{D}\frac{1}{%
\left( x-y\right) ^{2}},  \nonumber \\
\left( \gamma _{\mu }\otimes \lambda _{I}\right) _{A}^{B} &=&\left( \gamma
_{\mu }\right) _{\alpha }^{\beta }\left( \lambda _{I}\right) _{a}^{b}\delta
_{i}^{j},\quad \sum\limits_{I}\left( \lambda _{I}\right) _{a}^{b}\left(
\lambda _{I}\right) _{c}^{d}=\frac{1}{2}\left[ \delta _{c}^{b}\delta
_{a}^{d}-\frac{1}{N_{c}}\delta _{a}^{b}\delta _{c}^{d}\right] ,\quad
\label{43}
\end{eqnarray}
$\lambda _{I\text{ }}$ are the Gell-Mann matrices of the color symmetry
group. \noindent In order to describe the diquark systems we introduce some
composite bi-local bi-spinor fields $\Phi _{BD}\left( x,y\right) ,$%
\linebreak $\,\overline{\Phi }^{AC}\left( x,y\right) $ obeying the
Fermi-Dirac statistics 
\begin{equation}
\Phi _{DB}\left( y,x\right) =-\Phi _{BD}\left( x,y\right) ,\,\,\,\,\,%
\overline{\Phi }^{CA}\left( y,x\right) =-\overline{\Phi }^{AC}\left(
x,y\right) ,  \label{44}
\end{equation}
and the functional integral over these bosonic fields 
\begin{equation}
Z_{0}^{\Phi }=\int \left[ D\Phi \right] \left[ D\overline{\Phi }\right] \exp
\left\{ -\frac{1}{2}\int dx\int dy\overline{\Phi }^{AC}\left( x,y\right)
V_{CA}^{BD}\left( x-y\right) \Phi _{DB}\left( y,x\right) \right\} .
\label{45}
\end{equation}
By means the shift of the functional integration variables 
\begin{eqnarray*}
\Phi _{DB}\left( y,x\right) &\rightarrow &\Phi _{DB}\left( y,x\right) +\psi
_{D}\left( y\right) \psi _{B}\left( x\right) \\
\overline{\Phi }^{AC}\left( x,y\right) &\rightarrow &\overline{\Phi }%
^{AC}\left( x,y\right) +\overline{\psi }^{A}\left( x\right) \overline{\psi }%
^{C}\left( y\right) ,
\end{eqnarray*}
we can establish the Hubbard-Stratonovich transformation 
\begin{eqnarray}
&&\exp \left\{ \frac{1}{2}\int dx\int dy\overline{\psi }^{A}\left( x\right) 
\overline{\psi }^{C}\left( y\right) V_{CA}^{BD}\left( x-y\right) \psi
_{D}\left( y\right) \psi _{B}\left( x\right) \right\}  \label{46} \\
&=&\frac{1}{Z_{0}^{\Phi }}\int \left[ D\Phi \right] \left[ D\overline{\Phi }%
\right] \exp \left\{ -\frac{1}{2}\int dx\int dy\overline{\Phi }^{AC}\left(
x,y\right) V_{CA}^{BD}\left( x-y\right) \Phi _{DB}\left( y,x\right) \right\}
{}  \nonumber \\
&&\exp \left\{ -\frac{1}{2}\int dx\int dy\left[ \overline{\Delta }%
^{BD}\left( x,y\right) \psi _{D}\left( y\right) \psi _{B}\left( x\right) +%
\overline{\psi }^{A}\left( x\right) \overline{\psi }^{C}\left( y\right)
\Delta _{CA}\left( y,x\right) \right] \right\} ,  \nonumber
\end{eqnarray}
\begin{equation}
\QDATOP{\Delta _{CA}\left( y,x\right) =V_{CA}^{BD}\left( x-y\right) \Phi
_{DB}\left( y,x\right) ,}{\overline{\Delta }^{BD}\left( x,y\right) =%
\overline{\Phi }^{AC}\left( x,y\right) V_{CA}^{BD}\left( x-y\right) ,}
\label{47}
\end{equation}
\begin{equation}
\Delta _{AC}\left( x,y\right) =-\Delta _{CA}\left( y,x\right) ,\qquad 
\overline{\Delta }^{DB}\left( y,x\right) =-\overline{\Delta }^{BD}\left(
x,y\right) ,  \label{48}
\end{equation}
and rewrite the partition function 
\begin{eqnarray}
Z &=&\frac{1}{Z_{0}^{\Phi }}\int \left[ D\Phi \right] \left[ D\overline{\Phi 
}\right] \exp \left\{ -\frac{1}{2}\int dx\int dy\overline{\Phi }^{AC}\left(
x,y\right) V_{CA}^{BD}\left( x-y\right) \Phi _{DB}\left( y,x\right) \right\}
\nonumber \\
&&\int \left[ D\psi \right] \left[ D\overline{\psi }\right] \exp \left\{
-\int dx\overline{\psi }^{A}\left( x\right) L_{A}^{B}\psi _{B}\left(
x\right) \right\}  \label{49} \\
&&\exp \left\{ -\frac{1}{2}\int dx\int dy\left[ \overline{\Delta }%
^{BD}\left( x,y\right) \psi _{D}\left( y\right) \psi _{B}\left( x\right) +%
\overline{\psi }^{A}\left( x\right) \overline{\psi }^{C}\left( y\right)
\Delta _{CA}\left( y,x\right) \right] \right\}  \nonumber
\end{eqnarray}
in the form (23) with the effective action

\begin{equation}
S_{\text{eff}}\left[ \Phi ,\,\overline{\Phi }\right] =-\frac{1}{2}\int
dx\int dy\overline{\Phi }^{AC}\left( x,y\right) V_{CA}^{BD}\left( x-y\right)
\Phi _{DB}\left( y,x\right) +W\left[ \Delta ,\,\overline{\Delta }\right] ,
\label{50}
\end{equation}

\begin{eqnarray}
&&\exp \left\{ W\left[ \Delta ,\,\overline{\Delta }\right] \right\} = 
\nonumber \\
&=&1+\sum_{n=1}^{\infty }\Gamma ^{\left( 2n\right) }\left[ \Delta ,\,%
\overline{\Delta }\right]  \label{51} \\
&=&\left\langle \text{T}\left[ \exp \left\{ -\frac{1}{2}\int dx\int dy\left[ 
\overline{\Delta }^{BD}\left( x,y\right) \psi _{D}\left( y\right) \psi
_{B}\left( x\right) +\overline{\psi }^{A}\left( x\right) \overline{\psi }%
^{C}\left( y\right) \Delta _{CA}\left( y,x\right) \right] \right\} \right]
\right\rangle .  \nonumber
\end{eqnarray}
Calculations give 
\begin{eqnarray}
\Gamma ^{\left( 2\right) }\left[ \Delta ,\,\overline{\Delta }\right]
&=&W^{\left( 2\right) }\left[ \Delta ,\,\overline{\Delta }\right] =\frac{1}{2%
}\int dx_{1}\int dy_{1}\int dx_{2}\int dy_{2}  \label{52} \\
&&\qquad \overline{\Delta }^{A_{1}C_{1}}\left( x_{1},y_{1}\right)
S_{C_{1}}^{C_{2}}\left( y_{1}-y_{2}\right) \Delta _{C_{2}A_{2}}\left(
y_{2},x_{2}\right) S_{A_{1}}^{\text{T\thinspace }A_{2}}\left(
x_{2}-x_{1}\right) ,  \nonumber
\end{eqnarray}
\begin{equation}
S_{A_{1}}^{\text{T\thinspace }A_{2}}\left( x_{2}-x_{1}\right)
=S_{A_{1}}^{A_{2}}\left( x_{1}-x_{2}\right) ,  \label{53}
\end{equation}
\begin{equation}
\Gamma ^{\left( 4\right) }\left[ \Delta ,\,\overline{\Delta }\right] =\frac{1%
}{2}\left( W^{\left( 2\right) }\left[ \Delta ,\,\overline{\Delta }\right]
\right) ^{2}+W^{\left( 4\right) }\left[ \Delta ,\,\overline{\Delta }\right] ,
\label{54}
\end{equation}
\begin{eqnarray}
W^{\left( 4\right) }\left[ \Delta ,\,\overline{\Delta }\right] &=&-\frac{1}{4%
}\int dx_{1}\int dy_{1}\int dx_{2}\int dy_{2}\int dx_{3}\int dy_{3}\int
dx_{4}\int dy_{4}  \nonumber \\
&&\overline{\Delta }^{A_{1}C_{1}}\left( x_{1},y_{1}\right)
S_{C_{1}}^{C_{2}}\left( y_{1}-y_{2}\right) \Delta _{C_{2}A_{2}}\left(
y_{2},x_{2}\right) S_{A_{3}}^{\text{T\thinspace }A_{2}}\left(
x_{2}-x_{3}\right)  \hskip 0.8cm \label{55} \\
&&\overline{\Delta }^{A_{3}C_{3}}\left( x_{3},y_{3}\right)
S_{C_{3}}^{C_{4}}\left( y_{3}-y_{4}\right) \Delta _{C_{4}A_{4}}\left(
y_{4},x_{4}\right) S_{A_{1}}^{\text{T\thinspace }A_{4}}\left(
x_{4}-x_{1}\right) ,  \nonumber
\end{eqnarray}

\begin{equation}
\Gamma ^{\left( 6\right) }\left[ \Delta ,\,\overline{\Delta }\right] =\frac{1%
}{3!}\left( W^{\left( 2\right) }\left[ \Delta ,\,\overline{\Delta }\right]
\right) ^{3}+W^{\left( 2\right) }\left[ \Delta ,\,\overline{\Delta }\right]
W^{\left( 4\right) }\left[ \Delta ,\,\overline{\Delta }\right] +W^{\left(
6\right) }\left[ \Delta ,\,\overline{\Delta }\right] ,  \label{56}
\end{equation}
\begin{eqnarray}
W^{\left( 6\right) }\left[ \Delta ,\,\overline{\Delta }\right] &=&\frac{1}{6}%
\int dx_{1}\int dy_{1}\int dx_{2}\int dy_{2}\,...\,\int dx_{5}\int
dy_{5}\int dx_{6}\int dy_{6}  \nonumber \\
&&\overline{\Delta }^{A_{1}C_{1}}\left( x_{1},y_{1}\right)
S_{C_{1}}^{C_{2}}\left( y_{1}-y_{2}\right) \Delta _{C_{2}A_{2}}\left(
y_{2},x_{2}\right) S_{A_{3}}^{\text{T\thinspace }A_{2}}\left(
x_{2}-x_{3}\right)  \nonumber \\
&&\overline{\Delta }^{A_{3}C_{3}}\left( x_{3},y_{3}\right)
S_{C_{3}}^{C_{4}}\left( y_{3}-y_{4}\right) \Delta _{C_{4}A_{4}}\left(
y_{4},x_{4}\right) S_{A_{5}}^{\text{T\thinspace }A_{4}}\left(
x_{4}-x_{5}\right)  \hskip 0.8cm \label{57} \\
&&\overline{\Delta }^{A_{5}C_{5}}\left( x_{5},y_{5}\right)
S_{C_{5}}^{C_{6}}\left( y_{5}-y_{6}\right) \Delta _{C_{6}A_{6}}\left(
y_{6},x_{6}\right) S_{A_{1}}^{\text{T\thinspace }A_{6}}\left(
x_{6}-x_{1}\right) ,  \nonumber
\end{eqnarray}
\begin{eqnarray}
\Gamma ^{\left( 8\right) }\left[ \Delta ,\,\overline{\Delta }\right] &=&%
\frac{1}{4!}\left( W^{\left( 2\right) }\left[ \Delta ,\,\overline{\Delta }%
\right] \right) ^{4}+\frac{1}{2}\left( W^{\left( 2\right) }\left[ \Delta ,\,%
\overline{\Delta }\right] \right) ^{2}W^{\left( 4\right) }\left[ \Delta ,\,%
\overline{\Delta }\right]  \label{58} \\
&&+\frac{1}{2}\left( W^{\left( 4\right) }\left[ \Delta ,\,\overline{\Delta }%
\right] \right) ^{2}+W^{\left( 2\right) }\left[ \Delta ,\,\overline{\Delta }%
\right] W^{\left( 6\right) }\left[ \Delta ,\,\overline{\Delta }\right]
+W^{\left( 8\right) }\left[ \Delta ,\,\overline{\Delta }\right] ,  \nonumber
\end{eqnarray}
\begin{eqnarray}
W^{\left( 8\right) }\left[ \Delta ,\,\overline{\Delta }\right] &=&-\frac{1}{8%
}\int dx_{1}\int dy_{1}\int dx_{2}\int dy_{2}\,...\,\int dx_{7}\int
dy_{7}\int dx_{8}\int dy_{8}  \nonumber \\
&&\overline{\Delta }^{A_{1}C_{1}}\left( x_{1},y_{1}\right)
S_{C_{1}}^{C_{2}}\left( y_{1}-y_{2}\right) \Delta _{C_{2}A_{2}}\left(
y_{2},x_{2}\right) S_{A_{3}}^{\text{T\thinspace }A_{2}}\left(
x_{2}-x_{3}\right)  \nonumber \\
&&\overline{\Delta }^{A_{3}C_{3}}\left( x_{3},y_{3}\right)
S_{C_{3}}^{C_{4}}\left( y_{3}-y_{4}\right) \Delta _{C_{4}A_{4}}\left(
y_{4},x_{4}\right) S_{A_{5}}^{\text{T\thinspace }A_{4}}\left(
x_{4}-x_{5}\right)  \hskip 0.8cm \label{59} \\
&&\overline{\Delta }^{A_{5}C_{5}}\left( x_{5},y_{5}\right)
S_{C_{5}}^{C_{6}}\left( y_{5}-y_{6}\right) \Delta _{C_{6}A_{6}}\left(
y_{6},x_{6}\right) S_{A_{7}}^{\text{T\thinspace }A_{6}}\left(
x_{6}-x_{7}\right)  \nonumber \\
&&\overline{\Delta }^{A_{7}C_{7}}\left( x_{7},y_{7}\right)
S_{C_{7}}^{C_{8}}\left( y_{7}-y_{8}\right) \Delta _{C_{8}A_{8}}\left(
y_{8},x_{8}\right) S_{A_{1}}^{\text{T\thinspace }A_{8}}\left(
x_{8}-x_{1}\right) ,  \nonumber \\
&&.............................\,\,.  \nonumber
\end{eqnarray}
\noindent It is easy to verify that 
\begin{equation}
W\left[ \Delta ,\,\overline{\Delta }\right] =\sum_{n=1}^{\infty }W^{\left(
2n\right) }\left[ \Delta ,\,\overline{\Delta }\right] .  \label{60}
\end{equation}

From the variational principle for the effective action we derive the field
equation 
\begin{equation}
\frac{1}{2}\Delta _{CA}\left( y,x\right) =V_{CA}^{BD}\left( x-y\right)
\sum_{n=1}^{\infty }\frac{\delta \,W^{\left( 2n\right) }\left[ \Delta ,\,%
\overline{\Delta }\right] }{\delta \overline{\Delta }^{BD}\left( x,y\right) }%
.  \label{61}
\end{equation}
It has the explicit form 
\begin{eqnarray}
\Delta _{CA}\left( y,x\right) &=&V_{CA}^{BD}\left( x-y\right) \left\{ \int
dx_{2}\int dy_{2}S_{D}^{C_{2}}\left( y-y_{2}\right) \Delta
_{C_{2}A_{2}}\left( y_{2},x_{2}\right) S_{B}^{\text{T\thinspace }%
A_{2}}\left( x_{2}-x\right) \right.  \nonumber \\
&&-\int dx_{2}\int dy_{2}...\,\int dx_{4}\int dy_{4}S_{D}^{C_{2}}\left(
y-y_{2}\right) \Delta _{C_{2}A_{2}}\left( y_{2},x_{2}\right) S_{A_{3}}^{%
\text{T\thinspace }A_{2}}\left( x_{2}-x_{3}\right)  \nonumber \\
&&\hspace{2cm}\overline{\Delta }^{A_{3}C_{3}}\left( x_{3},y_{3}\right)
S_{C_{3}}^{C_{4}}\left( y_{3}-y_{4}\right) \Delta _{C_{4}A_{4}}\left(
y_{4},x_{4}\right) S_{B}^{\text{T}\,A_{4}}\left( x_{4}-x\right)  \nonumber \\
&&+\int dx_{2}\int dy_{2}...\,\int dx_{6}\int dy_{6}S_{D}^{C_{2}}\left(
y-y_{2}\right) \Delta _{C_{2}A_{2}}\left( y_{2},x_{2}\right) S_{A_{3}}^{%
\text{T\thinspace }A_{2}}\left( x_{2}-x_{3}\right)  \nonumber \\
&&\hspace{2cm}\overline{\Delta }^{A_{3}C_{3}}\left( x_{3},y_{3}\right)
S_{C_{3}}^{C_{4}}\left( y_{3}-y_{4}\right) \Delta _{C_{4}A_{4}}\left(
y_{4},x_{4}\right) S_{A_{5}}^{\text{T}\,A_{4}}\left( x_{4}-x_{5}\right) 
\nonumber \\
&&\hspace{2cm}\overline{\Delta }^{A_{5}C_{5}}\left( x_{5},y_{5}\right)
S_{C_{5}}^{C_{6}}\left( y_{5}-y_{6}\right) \Delta _{C_{6}A_{6}}\left(
y_{6},x_{6}\right) S_{B}^{\text{T}\,A_{6}}\left( x_{6}-x\right)  \nonumber \\
&&-\int dx_{2}\int dy_{2}...\,\int dx_{8}\int dy_{8}S_{D}^{C_{2}}\left(
y-y_{2}\right) \Delta _{C_{2}A_{2}}\left( y_{2},x_{2}\right) S_{A_{3}}^{%
\text{T\thinspace }A_{2}}\left( x_{2}-x_{3}\right)  \nonumber \\
&&\hspace{1.5cm}...\,\,\overline{\Delta }^{A_{7}C_{7}}\left(
x_{7},y_{7}\right) S_{C_{7}}^{C_{8}}\left( y_{7}-y_{8}\right) \Delta
_{C_{8}A_{8}}\left( y_{8},x_{8}\right) S_{B}^{\text{T}\,A_{8}}\left(
x_{8}-x\right)  \nonumber \\
&&+\left. ...\,\qquad \qquad \qquad \qquad \qquad \qquad \qquad \qquad 
\QDATOP{\,}{\,}\right\} .  \label{62}
\end{eqnarray}
Considering the solutions of this equation in the special class of functions
depending only on the difference of the coordinates 
\begin{equation}
\Delta _{CA}\left( y,x\right) =\Delta _{CA}\left( y-x\right) ,\qquad 
\overline{\Delta }^{BD}\left( x,y\right) =\overline{\Delta }^{BD}\left(
x-y\right) ,  \label{63}
\end{equation}
performing the Fourier transformations 
\begin{eqnarray}
\Delta _{CA}\left( \mathbf{y}-\mathbf{x},\sigma -\tau \right) &=&\frac{1}{%
\beta }\sum_{m}e^{i\varepsilon _{m}\left( \sigma -\tau \right) }\frac{1}{%
\left( 2\pi \right) ^{3}}\int d\mathbf{p}e^{i\mathbf{p}\left( \mathbf{y}-%
\mathbf{x}\right) }\widetilde{\Delta }_{CA}\left( \mathbf{p},\varepsilon
_{m}\right) ,  \nonumber \\
\overline{\Delta }^{BD}\left( \mathbf{x}-\mathbf{y},\tau -\sigma \right) &=&%
\frac{1}{\beta }\sum_{m}e^{i\varepsilon _{m}\left( \tau -\sigma \right) }%
\frac{1}{\left( 2\pi \right) ^{3}}\int d\mathbf{p}e^{i\mathbf{p}\left( 
\mathbf{x}-\mathbf{y}\right) }\widetilde{\overline{\Delta }}^{\,BD}\left( 
\mathbf{p},\varepsilon _{m}\right) , \hskip 0.8cm  \label{64} \\
V_{CA}^{DB}\left( \mathbf{x}-\mathbf{y},\tau -\sigma \right) &=&\frac{1}{%
\beta }\sum_{m}e^{i\omega _{m}\left( \sigma -\tau \right) }\frac{1}{\left(
2\pi \right) ^{3}}\int d\mathbf{p}e^{i\mathbf{p}\left( \mathbf{y}-\mathbf{x}%
\right) }\widetilde{V}_{CA}^{DB}\left( \mathbf{p},\omega _{m}\right) , 
\nonumber
\end{eqnarray}

\[
\varepsilon =\left( 2m+1\right) \frac{\pi }{\beta },\qquad \omega _{m}=2m%
\frac{\pi }{\beta }, 
\]
and introducing matrices $\widetilde{\Delta }\left( \mathbf{p},\varepsilon
_{m}\right) $, $\widetilde{\overline{\Delta }}\left( \mathbf{p},\varepsilon
_{m}\right) $, $\widetilde{S}\left( \mathbf{p},\varepsilon _{m}\right) $, $%
\widetilde{S}^{\text{T}}\left( -\mathbf{p},-\varepsilon _{m}\right) $ with
the elements $\widetilde{\Delta }_{CA}\left( \mathbf{p},\varepsilon
_{m}\right) $, $\widetilde{\overline{\Delta }}^{\,\,BD}\left( \mathbf{p}%
,\varepsilon _{m}\right) $, $\widetilde{S}_{A}^{B}\left( \mathbf{p}%
,\varepsilon _{m}\right) $, $\widetilde{S}_{B}^{\text{T\thinspace }A}\left( -%
\mathbf{p},-\varepsilon _{m}\right) $, we derive the extended BCS gap
equation for the superconducting quark-quark pairing in\ QCD 
\begin{eqnarray}
\widetilde{\Delta }_{CA}\left( \mathbf{p},\varepsilon _{m}\right) &=&\frac{1%
}{\beta }\sum_{n}\frac{1}{\left( 2\pi \right) ^{3}}\int d\mathbf{q}%
\widetilde{V}_{CA}^{BD}\left( \mathbf{p}-\mathbf{q},\varepsilon
_{m}-\varepsilon _{n}\right)  \nonumber \\
&&\left[ \widetilde{S}\left( \mathbf{q},\varepsilon _{n}\right) \widetilde{%
\Delta }\left( \mathbf{q},\varepsilon _{n}\right) \widetilde{S}^{\text{T}%
}\left( -\mathbf{q},-\varepsilon _{n}\right) \right.  \nonumber \\
&&\left. \frac{1}{1+\widetilde{\overline{\Delta }}\left( \mathbf{q}%
,\varepsilon _{n}\right) \widetilde{S}\left( \mathbf{q},\varepsilon
_{n}\right) \widetilde{\Delta }\left( \mathbf{q},\varepsilon _{n}\right) 
\widetilde{S}^{\text{T}}\left( -\mathbf{q},-\varepsilon _{n}\right) }\right]
_{DB}.  \label{65}
\end{eqnarray}
At the values of the fields $\Delta _{CA}\left( y,x\right) $ and $\overline{%
\Delta }^{BD}\left( x,y\right) $ satisfying the equation (62) the effective
action (50) equals 
\begin{eqnarray}
S_{eff}\left[ \Phi ,\overline{\Phi }\right] &=&W\left[ \Delta ,\,\overline{%
\Delta }\right] -\int dx\int dy\overline{\Delta }^{AC}\left( x,y\right) 
\frac{\delta W\left[ \Delta ,\,\overline{\Delta }\right] }{\delta \overline{%
\Delta }^{AC}\left( x,y\right) }  \nonumber \\
&=&-\left( \frac{1}{4}-\frac{1}{2}\right) \int dx_{1}\int dy_{1}...\,\,\int
dx_{4}\int dy_{4}\limfunc{Tr}\left[ \overline{\Delta }\left(
x_{1},y_{1}\right) S\left( y_{1}-y_{2}\right) \Delta \left(
y_{2},x_{2}\right) \right.  \nonumber \\
&&\qquad \left. S^{\text{T}}\left( x_{2}-x_{3}\right) \overline{\Delta }%
\left( x_{3},y_{3}\right) S\left( y_{3}-y_{4}\right) \Delta \left(
y_{4},x_{4}\right) S^{\text{T}}\left( x_{4}-x_{1}\right) \right]  \nonumber
\\
&&+\left( \frac{1}{6}-\frac{1}{2}\right) \int dx_{1}\int dy_{1}...\,\,\int
dx_{6}\int dy_{6}\limfunc{Tr}\left[ \overline{\Delta }\left(
x_{1},y_{1}\right) S\left( y_{1}-y_{2}\right) \Delta \left(
y_{2},x_{2}\right) \right.  \nonumber \\
&&\qquad \left. S^{\text{T}}\left( x_{2}-x_{3}\right) ...\,\,\overline{%
\Delta }\left( x_{5},y_{5}\right) S\left( y_{5}-y_{6}\right) \Delta \left(
y_{6},x_{6}\right) S^{\text{T}}\left( x_{6}-x_{1}\right) \right]  \nonumber
\\
&&-\left( \frac{1}{8}-\frac{1}{2}\right) \int dx_{1}\int dy_{1}...\,\,\int
dx_{8}\int dy_{8}\limfunc{Tr}\left[ \overline{\Delta }\left(
x_{1},y_{1}\right) S\left( y_{1}-y_{2}\right) \Delta \left(
y_{2},x_{2}\right) \right.  \nonumber \\
&&\qquad \left. S^{\text{T}}\left( x_{2}-x_{3}\right) ...\,\,\overline{%
\Delta }\left( x_{7},y_{7}\right) S\left( y_{7}-y_{8}\right) \Delta \left(
y_{8},x_{8}\right) S^{\text{T}}\left( x_{8}-x_{1}\right) \right]  \nonumber
\\
&&+\,...\,,  \label{66}
\end{eqnarray}
where $\Delta \left( y,x\right) $, $\overline{\Delta }\left( y,x\right) $,$%
\,S\left( x-y\right) $, $S^{\text{T}}\left( x-y\right) $ denote the matrices
with the elements $\Delta _{CA}\left( y,x\right) $, $\overline{\Delta }%
^{AC}\left( x,y\right) $, $S_{A}^{B}\left( x-y\right) $, $S_{B}^{\text{T}%
\,A}\left( x-y\right) $. In the case of the special class (63) of the
solutions $\Delta _{CA}\left( y-x\right) $, $\overline{\Delta }^{AC}\left(
x-y\right) $ the effective action is expressed in terms of the free energy
by means of the formula (37). Comparing the expressions (37) and (66) and
performing the Fourier transformation, we obtain

\begin{eqnarray}
F\left[ \mathbf{x},\Delta \right] &=&F\left[ \Delta \right] =-\frac{1}{\beta 
}\sum_{m}\frac{1}{\left( 2\pi \right) ^{3}}\int d\mathbf{p}  \nonumber \\
&&\limfunc{Tr}\left\{ \left( \frac{1}{2}-\frac{1}{4}\right) \left[ 
\widetilde{\overline{\Delta }}\left( \mathbf{p},\varepsilon _{m}\right) 
\widetilde{S}\left( \mathbf{p},\varepsilon _{m}\right) \widetilde{\Delta }%
\left( \mathbf{p},\varepsilon _{m}\right) \widetilde{S}^{\text{T}}\left( -%
\mathbf{p},-\varepsilon _{m}\right) \right] ^{2}\right.  \nonumber \\
&&-\left( \frac{1}{2}-\frac{1}{6}\right) \left[ \widetilde{\overline{\Delta }%
}\left( \mathbf{p},\varepsilon _{m}\right) \widetilde{S}\left( \mathbf{p}%
,\varepsilon _{m}\right) \widetilde{\Delta }\left( \mathbf{p},\varepsilon
_{m}\right) \widetilde{S}^{\text{T}}\left( -\mathbf{p},-\varepsilon
_{m}\right) \right] ^{3}  \nonumber \\
&&+\left( \frac{1}{2}-\frac{1}{8}\right) \left[ \widetilde{\overline{\Delta }%
}\left( \mathbf{p},\varepsilon _{m}\right) \widetilde{S}\left( \mathbf{p}%
,\varepsilon _{m}\right) \widetilde{\Delta }\left( \mathbf{p},\varepsilon
_{m}\right) \widetilde{S}^{\text{T}}\left( -\mathbf{p},-\varepsilon
_{m}\right) \right] ^{4}  \label{67} \\
&&\left. ....\qquad \qquad \qquad \qquad \qquad \qquad \hspace{0in}\hspace{%
0in}\hspace{0in}\hspace{0in}\hspace{0in}\hspace{0in}\hspace{0in}\hspace{0in}%
\hspace{0in}\hspace{0in}\hspace{0in}\hspace{1in}.. 
\begin{array}{l}
\end{array}
\right\} .  \nonumber
\end{eqnarray}
This series can be also written in the compact form of the integral
representation
\begin{eqnarray}
F\left[ \Delta \right] &=&\frac{1}{\beta }\sum_{m}\frac{1}{\left( 2\pi
\right) ^{3}}\int d\mathbf{p}\frac{1}{2}\limfunc{Tr}\left[ \widetilde{S}%
\left( \mathbf{p},\varepsilon _{m}\right) \Delta \left( \mathbf{p}%
,\varepsilon _{m}\right) \widetilde{S}^{\text{T}}\left( -\mathbf{p}%
,-\varepsilon _{m}\right) \overline{\Delta }\left( \mathbf{p},\varepsilon
_{m}\right) \right.  \nonumber \\
&&\left\{ \left. \frac{1}{1+\widetilde{S}\left( \mathbf{p},\varepsilon
_{m}\right) \Delta \left( \mathbf{p},\varepsilon _{m}\right) \widetilde{S}^{%
\text{T}}\left( -\mathbf{p},-\varepsilon _{m}\right) \overline{\Delta }%
\left( \mathbf{p},\varepsilon _{m}\right) }\right. \right.  \nonumber \\
&&-\left. \left. \int\limits_{0}^{1}d\alpha \frac{1}{1+\alpha \widetilde{S}%
\left( \mathbf{p},\varepsilon _{m}\right) \Delta \left( \mathbf{p}%
,\varepsilon _{m}\right) \widetilde{S}^{\text{T}}\left( -\mathbf{p}%
,-\varepsilon _{m}\right) \overline{\Delta }\left( \mathbf{p},\varepsilon
_{m}\right) }\right\} \right] .  \label{68}
\end{eqnarray}

In order to simplify the expression for $\Delta _{AC}\left( \mathbf{p}%
,\varepsilon _{m}\right) $ we introduce 4-vector $p_{\mu }$ with the fourth
component $p_{4}=\varepsilon _{m}$ and denote $\Delta _{AC}\left( \mathbf{p}%
,\varepsilon _{m}\right) $by $\Delta _{AC}\left( p\right) $. The condition
(48) is now written in the form 
\begin{equation}
\Delta _{BA}\left( -p\right) =-\Delta _{AB}\left( p\right) .  \label{69}
\end{equation}
The bi-spinors $\Delta _{AB}\left( p\right) $ are expressed in terms of the
Dirac bi-spinors 
\begin{eqnarray}
\Delta _{AB}\left( p\right) &=&\Delta _{\left( ai\alpha \right) \left(
bj\beta \right) }\left( p\right) =\left( \gamma _{5}C\right) _{\alpha \beta
}\Delta _{\left( ai\right) \left( bj\right) }^{S}\left( p\right) +\left(
C\right) _{\alpha \beta }\Delta _{\left( ai\right) \left( bj\right)
}^{P}\left( p\right)  \nonumber \\
&&+\left( \gamma _{\mu }\gamma _{5}C\right) _{\alpha \beta }\Delta _{\mu
\,\left( ai\right) \left( bj\right) }^{V}\left( p\right) +\left( \gamma
_{\mu }C\right) _{\alpha \beta }\Delta _{\mu \,\left( ai\right) \left(
bj\right) }^{A}\left( p\right)  \label{70} \\
&&+\left( \sigma _{\mu \nu }\gamma _{5}C\right) _{\alpha \beta }\,\Delta
_{\mu \nu \,\left( ai\right) \left( bj\right) }^{t}\left( p\right) . 
\nonumber
\end{eqnarray} 
From the conditions (69) it follows that under the interchange $\left(
ai\right) \leftrightarrow \left( bj\right) $ and the inversion $p\rightarrow
-p$ the scalar, pseudoscalar and vector functions are invariant while the
pseudovector and tensor functions change their sign 
\begin{eqnarray}
\QATOP{\Delta _{\left( bj\right) \left( ai\right) }^{S,\,P}\left( -p\right)
=\Delta _{\left( ai\right) \left( bj\right) }^{S,\,P}\left( p\right) ,\qquad
\Delta _{\mu \,\left( bj\right) \left( ai\right) }^{V}\left( -p\right)
=\Delta _{\mu \,\left( ai\right) \left( bj\right) }^{V}\left( p\right) ,} \nonumber \\
{\Delta _{\mu \,\left( bj\right) \left( ai\right) }^{A}\left( -p\right)
=-\Delta _{\mu \,\left( ai\right) \left( bj\right) }^{A}\left( p\right)
,\qquad \Delta _{\mu \nu \,\left( bj\right) \left( ai\right) }^{t}\left(
-p\right) =-\Delta _{\mu \nu \,\left( ai\right) \left( bj\right) }^{t}\left(
p\right) .}  \label{71}
\end{eqnarray}

In general, the existence of non-vanishing superconducting order parameters
which are not the singlets of the color and/or flavor groups and lower the
free energy would mean the spontaneous breaking of the color and/or flavor
symmetries. For the systems with the isomorphic color and flavor groups $%
SU\left( N\right) _{c}$ and $SU\left( N\right) _{f}$ there may exist the
superconducting order parameters which are the irreducible spinor
representations of the groups $SU\left( N\right) _{c}$ and $SU\left(
N\right) _{f}$ but the singlet of the ''diagonal'' $SU\left( N\right) $
subgroup of the direct product $SU\left( N\right) _{c}\otimes SU\left(
N\right) _{f}$. In this case we have the ''color-flavor locking''.

\section{Quark-Antiquark Pairing.}

The direct four-fermion coupling of the quark fields with the interaction
Lagrangian (16) or the non-local interaction of the quark fields with the
effective action given in the r.h.s of the formula (42) are also the origins
of the quark-antiquark pairing. In the case of the quark-antiquark pairing
due to the direct four-fermion coupling of the quark field we use the
interaction Lagrangian in the form

\begin{eqnarray}
L_{int} &=&\frac{1}{2}\overline{\psi }^{A}\left( x\right) \psi _{B}\left(
x\right) U_{AC}^{BD}\overline{\psi }^{C}\left( x\right) \psi _{D}\left(
x\right) ,  \nonumber \\
U_{CA}^{DB} &=&-U_{AC}^{BD}=-U_{CA}^{DB}=U_{AC}^{DB},  \label{72}
\end{eqnarray}
where instead of the constants $V_{CA}^{BD}$ in (16) we use the new notations

\begin{equation}
U_{AC}^{BD}=V_{CA}^{BD}.
\end{equation}
The partition function of the system equals 
\begin{eqnarray}
Z &=&\int \left[ D\psi \right] \left[ D\overline{\psi }\right] \exp \left\{
-\int dx\overline{\psi }^{A}\left( x\right) L_{A}^{B}\psi _{B}\left(
x\right) \right\}  \nonumber \\
&&\exp \left\{ \frac{1}{2}\int dx\overline{\psi }^{A}\left( x\right) \psi
_{B}\left( x\right) U_{AC}^{BD}\overline{\psi }^{C}\left( x\right) \psi
_{D}\left( x\right) \right\} .  \label{74}
\end{eqnarray}
Introducing the local hermitian fields $\Phi _{A}^{B}\left( x\right) $ and
the functional integral 
\begin{equation}
Z_{0}^{\Phi }=\int \left[ D\Phi \right] \exp \left\{ -\frac{1}{2}\int dx\Phi
_{B}^{A}\left( x\right) U_{AC}^{BD}\Phi _{D}^{C}\left( x\right) \right\} ,
\label{75}
\end{equation}
we establish the Hubbard-Stratonovich transformation 
\begin{eqnarray}
&&\exp \left\{ \frac{1}{2}\int dx\overline{\psi }^{A}\left( x\right) \psi
_{B}\left( x\right) U_{AC}^{BD}\overline{\psi }^{C}\left( x\right) \psi
_{D}\left( x\right) \right\} =  \nonumber \\
&=&\frac{1}{Z_{0}^{\Phi }}\int \left[ D\Phi \right] \exp \left\{ -\frac{1}{2}%
\int dx\Phi _{B}^{A}\left( x\right) U_{AC}^{BD}\Phi _{D}^{C}\left( x\right)
\right\}  \nonumber \\
&&\exp \left\{ -\int dx\overline{\psi }^{A}\left( x\right) \psi _{B}\left(
x\right) \Delta _{A}^{B}\left( x\right) \right\} ,\qquad  \label{76}
\end{eqnarray}
\begin{equation}
\Delta _{A}^{B}\left( x\right) =U_{AC}^{BD}\Phi _{D}^{C}\left( x\right) ,
\label{77}
\end{equation}
and rewrite the partition function in the form

\begin{equation}
Z=\frac{Z_{0}}{Z_{0}^{\Phi }}\int \left[ D\Phi \right] \exp \left\{
S_{eff}\left[ \Phi \right] \right\}
\end{equation}
with the effective action 
\begin{equation}
S_{eff}\left[ \Phi \right] =-\frac{1}{2}\int dx\Phi _{B}^{A}\left( x\right)
U_{AC}^{BD}\Phi _{D}^{C}\left( x\right) +W\left[ \Delta \right] ,  \label{79}
\end{equation}
\begin{equation}
\exp \left\{ W\left[ \Delta \right] \right\} =1+\sum_{n=1}^{\infty }\Gamma
^{\left( n\right) }\left[ \Delta \right] =\left\langle \text{T}\left[ \exp
\left\{ -\int dx\Delta _{A}^{B}\left( x\right) \overline{\psi }^{A}\left(
x\right) \psi _{B}\left( x\right) \right\} \right] \right\rangle .
\label{80}
\end{equation}
Calculations give 
\begin{equation}
W\left[ \Delta \right] =\sum_{n=1}^{\infty }W^{\left( n\right) }\left[
\Delta \right] ,  \label{81}
\end{equation}
\begin{equation}
W^{\left( 1\right) }\left[ \Delta \right] =\int dx\Delta _{A}^{B}\left(
x\right) S_{B}^{A}\left( 0\right) ,  \label{82}
\end{equation}
\begin{equation}
W^{\left( 2\right) }\left[ \Delta \right] =-\frac{1}{2}\int dx_{1}\int
dx_{2}\,\Delta _{A_{1}}^{B_{1}}\left( x_{1}\right) S_{B_{1}}^{A_{2}}\left(
x_{1}-x_{2}\right) \Delta _{A_{2}}^{B_{2}}\left( x_{2}\right)
S_{B_{2}}^{A_{1}}\left( x_{2}-x_{1}\right) ,  \label{83}
\end{equation}
\begin{eqnarray}
W^{\left( 3\right) }\left[ \Delta \right] &=&\frac{1}{3}\int dx_{1}...\,\int
dx_{3}\,\Delta _{A_{1}}^{B_{1}}\left( x_{1}\right) S_{B_{1}}^{A_{2}}\left(
x_{1}-x_{2}\right) \Delta _{A_{2}}^{B_{2}}\left( x_{2}\right)  \nonumber \\
&&S_{B_{2}}^{A_{3}}\left( x_{2}-x_{3}\right) \Delta _{A_{3}}^{B_{3}}\left(
x_{3}\right) S_{B_{3}}^{A_{1}}\left( x_{3}-x_{1}\right) ,  \label{84}
\end{eqnarray}
...... 
\begin{eqnarray}
W^{\left( n\right) }\left[ \Delta \right] &=&\frac{\left( -1\right) ^{n+1}}{n%
}\int dx_{1}...\,\int dx_{n}\Delta _{A_{1}}^{B_{1}}\left( x_{1}\right)
S_{B_{1}}^{A_{2}}\left( x_{1}-x_{2}\right) \Delta _{A_{2}}^{B_{2}}\left(
x_{2}\right)  \nonumber \\
&&S_{B_{2}}^{A_{3}}\left( x_{2}-x_{3}\right) ...\,\Delta
_{A_{n}}^{B_{n}}\left( x_{n}\right) S_{B_{n}}^{A_{1}}\left(
x_{n}-x_{1}\right) .  \label{85}
\end{eqnarray}

From the variational principle 
\begin{equation}
\frac{\delta \,S_{eff}\left( \Phi \right) }{\delta \,\Phi _{D}^{C}\left(
x\right) }=0  \label{86}
\end{equation}
we derive the field equation 
\begin{equation}
\Delta _{C}^{D}\left( x\right) =U_{CA}^{DB}\frac{\delta W\left[ \Delta
\right] }{\delta \Delta _{A}^{B}\left( x\right) }=U_{CA}^{DB}\mathbf{S}%
_{B}^{A}\left( x,x\right) ,  \label{87}
\end{equation}
where $\mathbf{S}_{B}^{A}\left( y,x\right) \,$is the two-point Green
function of the quark field in the presence of the pairing interaction 
\begin{eqnarray}
\mathbf{S}_{B}^{A}\left( y,x\right) &=&S_{B}^{A}\left( y-x\right) -\int
dx_{1}S_{B}^{A_{1}}\left( y-x_{1}\right) \Delta _{A_{1}}^{B_{1}}\left(
x_{1}\right) S_{B_{1}}^{A}\left( x_{1}-x\right)  \nonumber \\
&&+\int dx_{1}\int dx_{2}S_{B}^{A_{1}}\left( y-x_{1}\right) \Delta
_{A_{1}}^{B_{1}}\left( x_{1}\right) S_{B_{1}}^{A_{2}}\left(
x_{1}-x_{2}\right) \Delta _{A_{2}}^{B_{2}}\left( x_{2}\right)
S_{B_{2}}^{A}\left( x_{2}-x\right)  \nonumber \\
&&-\int dx_{1}...\,\int dx_{3}S_{B}^{A_{1}}\left( y-x_{1}\right) \Delta
_{A_{1}}^{B_{1}}\left( x_{1}\right) S_{B_{1}}^{A_{2}}\left(
x_{1}-x_{2}\right) \Delta _{A_{2}}^{B_{2}}\left( x_{2}\right)  \label{88} \\
&&\qquad S_{B_{2}}^{A_{3}}\left( x_{2}-x_{3}\right) \Delta
_{A_{3}}^{B_{3}}\left( x_{3}\right) S_{B_{3}}^{A}\left( x_{3}-x\right)
+...\,.  \nonumber
\end{eqnarray}
It satisfies the Schwinger-Dyson equation 
\begin{equation}
\mathbf{S}_{B}^{A}\left( y,x\right) =S_{B}^{A}\left( y-x\right) -\int
dzS_{B}^{C}\left( y-z\right) \Delta _{C}^{D}\left( z\right) \mathbf{S}%
_{D}^{A}\left( z,x\right) .  \label{89}
\end{equation}
In the special class of the constant solutions of the field equation (87) 
\begin{equation}
\Delta _{B}^{A}\left( x\right) =\Delta _{B}^{\,A}=\limfunc{const}  \label{90}
\end{equation}
$\mathbf{S}_{B}^{A}\left( y,x\right) $ depends only on the coordinate
difference 
\begin{equation}
\mathbf{S}_{B}^{A}\left( y,x\right) =\mathbf{S}_{B}^{A}\left( y-x\right) .
\label{91}
\end{equation}
For its Fourier transform we have then an algebraic equation. Denote $\Delta 
$ the matrix with the elements $\Delta _{A}^{B}$. From the equation (89) it
follows that 
\begin{equation}
\widetilde{\mathbf{S}}\left( \mathbf{p},\varepsilon _{m}\right) =\widetilde{S%
}\left( \mathbf{p},\varepsilon _{m}\right) -\widetilde{S}\left( \mathbf{p}%
,\varepsilon _{m}\right) \Delta \widetilde{\mathbf{S}}\left( \mathbf{p}%
,\varepsilon _{m}\right)  \label{92}
\end{equation}
and 
\begin{equation}
\frac{1}{\widetilde{\mathbf{S}}\left( \mathbf{p},\varepsilon _{m}\right) }=%
\frac{1}{\widetilde{S}\left( \mathbf{p},\varepsilon _{m}\right) }+\Delta .
\label{93}
\end{equation}

With the field satisfying the equation (87) the effective action equals 
\begin{eqnarray}
S_{eff}\left[ \Phi \right] &=&W\left[ \Delta \right] -\frac{1}{2}\int
dx\Delta _{A}^{B}\left( x\right) \frac{\delta W\left[ \Delta \right] }{%
\delta \Delta _{A}^{B}\left( x\right) }=\left( 1-\frac{1}{2}\right) \int
dx\Delta _{A}^{B}\left( x\right) S_{B}^{A}\left( 0\right)  \nonumber \\
&&+\left( \frac{1}{3}-\frac{1}{2}\right) \int dx_{1}\int dx_{2}\,\int
dx_{3}\,\Delta _{A_{1}}^{B_{1}}\left( x_{1}\right) S_{B_{1}}^{A_{2}}\left(
x_{1}-x_{2}\right) \Delta _{A_{2}}^{B_{2}}\left( x_{2}\right)  \nonumber \\
&&\,\,\,\qquad S_{B_{2}}^{A_{3}}\left( x_{2}-x_{3}\right) \Delta
_{A_{3}}^{B_{3}}\left( x_{3}\right) S_{B_{3}}^{A_{1}}\left(
x_{3}-x_{1}\right)  \label{94} \\
&&-\left( \frac{1}{4}-\frac{1}{2}\right) \int dx_{1}...\,\,\int
dx_{4}\,\Delta _{A_{1}}^{B_{1}}\left( x_{1}\right) S_{B_{1}}^{A_{2}}\left(
x_{1}-x_{2}\right) \Delta _{A_{2}}^{B_{2}}\left( x_{2}\right)  \nonumber \\
&&\,\,\,\qquad S_{B_{2}}^{A_{3}}\left( x_{2}-x_{3}\right) ...\,\Delta
_{A_{4}}^{B_{4}}\left( x_{4}\right) S_{B_{4}}^{A_{1}}\left(
x_{4}-x_{1}\right) +...\,.  \nonumber \\
&=&\limfunc{Tr}\left\{ \left( 1-\frac{1}{2}\right) \int dx\Delta \left(
x\right) S\left( 0\right) \right.  \nonumber \\
&&+\left( \frac{1}{3}-\frac{1}{2}\right) \int dx_{1}\int dx_{2}\,\int
dx_{3}\,\Delta \left( x_{1}\right) S\left( x_{1}-x_{2}\right)  \nonumber \\
&&\,\,\,\qquad \Delta \left( x_{2}\right) S\left( x_{2}-x_{3}\right) \Delta
\left( x_{3}\right) S\left( x_{3}-x_{1}\right)  \nonumber \\
&&-\left( \frac{1}{4}-\frac{1}{2}\right) \int dx_{1}...\,\,\int
dx_{4}\,\Delta \left( x_{1}\right) S\left( x_{1}-x_{2}\right) \Delta \left(
x_{2}\right)  \nonumber \\
&&\,\qquad \left. S\left( x_{2}-x_{3}\right) ...\,\Delta \left( x_{4}\right)
S\left( x_{4}-x_{1}\right) +...\,.\hspace{1.7in}\QDATOP{\,}{\,}\right\} , 
\nonumber
\end{eqnarray}
where $\Delta \left( x\right) $ is the matrix with elements $\Delta
_{A}^{B}\left( x\right) $. It follows that in the case of the constant
solutions (90) of the field equations (87) we have following formula
determining the free energy density $F\left[ \mathbf{x};\Delta \right] $: 
\begin{eqnarray}
F\left[ \mathbf{x};\Delta \right] &=&F\left[ \Delta \right] =\left( 1-\frac{1%
}{2}\right) \Delta _{A}^{B}\,S_{B}^{A}\left( 0\right)  \nonumber \\
&&+\left( \frac{1}{3}-\frac{1}{2}\right) \int dy\int dz\Delta
_{A}^{B}\,S_{B}^{C}\left( x-y\right) \Delta _{C}^{D}\,S_{D}^{E}\left(
y-z\right) \Delta _{E}^{F}\,S_{F}^{A}\left( z-x\right)  \nonumber \\
&&-\left( \frac{1}{4}-\frac{1}{2}\right) \int dy\int dz\int dw\,\Delta
_{A}^{B}\,S_{B}^{C}\left( x-y\right) \Delta _{C}^{D}\,S_{D}^{E}\left(
y-z\right)  \nonumber \\
&&\qquad \Delta _{E}^{F}S_{F}^{G}\left( z-w\right) \Delta
_{G}^{H}S_{H}^{A}\left( w-x\right) +...  \nonumber \\
&=&\limfunc{Tr}\left\{ \left( 1-\frac{1}{2}\right) \Delta \,S\left( 0\right)
\right.  \label{95} \\
&&+\left( \frac{1}{3}-\frac{1}{2}\right) \Delta \int dy\int dzS\left(
x-y\right) \Delta \,S\left( y-z\right) \Delta \,S\left( z-x\right)  \nonumber
\\
&&-\left( \frac{1}{4}-\frac{1}{2}\right) \Delta \int dy\int dz\int
dw\,S\left( x-y\right) \Delta S\left( y-z\right)  \nonumber \\
&&\left. \Delta \,\,S\left( z-w\right) \Delta \,S\left( w-x\right) +...%
\hspace{2in}\QDATOP{\,}{\,}\right\} \qquad  \nonumber \\
&=&\frac{1}{\beta }\sum_{m}\frac{1}{\left( 2\pi \right) ^{3}}\int d\mathbf{p}%
\limfunc{Tr}\left\{ \left( 1-\frac{1}{2}\right) \Delta \,\widetilde{S}\left( 
\mathbf{p},\varepsilon _{m}\right) \right.  \nonumber \\
&&+\left. \left( \frac{1}{3}-\frac{1}{2}\right) \left[ \Delta \,\widetilde{S}%
\left( \mathbf{p},\varepsilon _{m}\right) \right] ^{3}-\left( \frac{1}{4}-%
\frac{1}{2}\right) \left[ \Delta \,\widetilde{S}\left( \mathbf{p}%
,\varepsilon _{m}\right) \right] ^{4}+...\,\right\} .  \nonumber
\end{eqnarray}
Summing up the infinite series, we obtain

\begin{equation}
F\left[ \Delta \right] =\frac{1}{\beta }\sum_{m}\frac{1}{\left( 2\pi \right)
^{3}}\int d\mathbf{p}\limfunc{Tr}\left\{ \widetilde{\Delta }\left( \mathbf{p}%
,\varepsilon _{m}\right) \left[ \int_{0}^{1}\widetilde{\mathbf{S}}^{\xi
}\left( \mathbf{p},\varepsilon _{m}\right) d\xi -\frac{1}{2}\widetilde{%
\mathbf{S}}\left( \mathbf{p},\varepsilon _{m}\right) \right] \right\} ,
\label{96}
\end{equation}
where $\widetilde{\mathbf{S}}\left( \mathbf{p},\varepsilon _{m}\right) $
satisfies the equation (93) and $\widetilde{\mathbf{S}}^{\xi }\left( \mathbf{%
p},\varepsilon _{m}\right) $ is determined by a similar one with the
replacement of $\Delta $ by $\xi \Delta $: 
\begin{equation}
\frac{1}{\widetilde{\mathbf{S}}^{\xi }\left( \mathbf{p},\varepsilon
_{m}\right) }=\frac{1}{\widetilde{S}\left( \mathbf{p},\varepsilon
_{m}\right) }+\xi \Delta .  \label{97}
\end{equation}
The order parameters $\Delta _{A}^{B}$ have the form 
\begin{equation}
\Delta _{A}^{B}=\Delta _{\left( ai\alpha \right) }^{\left( bj\beta \right)
}=\delta _{\alpha }^{\beta }\Delta _{\left( ai\right) }^{S\,\left( bj\right)
}\,+\left( \gamma _{5}\right) _{\alpha }^{\beta }\,\Delta _{\left( ai\right)
}^{P\,\left( bj\right) }.  \label{98}
\end{equation}
The non-vanishing order parameters $\Delta _{\left( ai\right) }^{P\,\left(
bj\right) }$ lowering the free energy would mean the spontaneous parity
conservation violation. If $M=0$ and the interaction Lagrangian (104) is
invariant under the chiral transformations, then the existence of
non-vanishing order parameters $\Delta _{\left( ai\right) }^{S\,\left(
bj\right) }$ and/or $\Delta _{\left( ai\right) }^{P\,\left( bj\right) }$
lowering the free energy would signify the spontaneous breaking of the
chiral invariance. If $\Delta _{\left( ai\right) }^{S\,\left( bj\right) }$
and/or $\Delta _{\left( ai\right) }^{P\,\left( bj\right) }$ are not the
singlets of the color and/or flavor group, then the color and/or flavor
symmetries are spontaneously broken. For the system with the isomorphic
color and flavor groups $SU\left( N\right) _{c}$ and $SU\left( N\right) _{f}$
there may exist the superconducting order parameters which are the
irreducible spinor representations of the groups $SU\left( N\right) _{c}$
and $SU\left( N\right) _{f}$ but the singlet of the the ''diagonal'' $%
SU\left( N\right) $ subgroup of the direct product $SU\left( N\right)
_{c}\otimes $ $SU\left( N\right) _{f}$. In this case we have the
''color-flavor locking''.

Now we consider the quark-antiquark pairing generated by the effective
non-local interaction of the quark fields due to the gluon exchange. For
this purpose we rewrite the partition function (42) in the appropriate form

\begin{eqnarray}
Z &=&\int \left[ D\psi \right] \left[ D\overline{\psi }\right] \exp \left\{
-\int dx\overline{\psi }^{A}\left( x\right) L_{A}^{B}\psi _{B}\left(
x\right) \right\}  \nonumber \\
&&\exp \left\{ \frac{1}{2}\int dx\int dy\overline{\psi }^{A}\left( x\right)
\psi _{B}\left( y\right) U_{AC}^{BD}\left( x-y\right) \overline{\psi }%
^{C}\left( y\right) \psi _{D}\left( x\right) \right\} .  \label{99}
\end{eqnarray}
with the new notations 
\begin{equation}
U_{AC}^{BD}\left( x-y\right) =-V_{AC}^{DB}\left( x-y\right) ,  \label{100}
\end{equation}
$V_{AC}^{DB}\left( x-y\right) $ being given in the formula (43). \noindent
Introduce the hermitian bi-local bosonic fields $\Phi _{B}^{A}\left(
x,y\right) $ and the functional integral 
\begin{equation}
Z_{0}^{\Phi }=\int \left[ D\Phi \right] \exp \left\{ -\frac{1}{2}\int dx\int
dy\Phi _{B}^{A}\left( x,y\right) U_{AC}^{BD}\left( x-y\right) \Phi
_{D}^{C}\left( y,x\right) \right\} .  \label{101}
\end{equation}
By shifting the functional integration variables, we obtain 
\begin{eqnarray}
&&\exp \left\{ \frac{1}{2}\int dx\int dy\overline{\psi }^{A}\left( x\right)
\psi _{B}\left( y\right) U_{AC}^{BD}\left( x-y\right) \overline{\psi }%
^{C}\left( y\right) \psi _{D}\left( x\right) \right\} =  \label{102} \\
&&\qquad =\frac{1}{Z_{0}^{\Phi }}\int \left[ D\Phi \right] \exp \left\{ -%
\frac{1}{2}\int dx\int dy\Phi _{B}^{A}\left( x,y\right) U_{AC}^{BD}\left(
x-y\right) \Phi _{D}^{C}\left( y,x\right) \right\}  \nonumber \\
&&\hspace{1.3in}\exp \left\{ -\int dx\int dy\overline{\psi }^{A}\left(
x\right) \psi _{B}\left( y\right) \Delta _{A}^{B}\left( x,y\right) \right\} ,
\nonumber
\end{eqnarray}
where 
\begin{equation}
\Delta _{A}^{B}\left( x,y\right) =U_{AC}^{BD}\left( x-y\right) \Phi
_{D}^{C}\left( y,x\right) ,  \label{103}
\end{equation}
and rewrite the partition function in the form (78) with the effective action

\begin{equation}
S_{\text{eff}}\left[ \Phi \right] =-\frac{1}{2}\int dx\int dy\Phi
_{B}^{A}\left( x,y\right) U_{AC}^{BD}\left( x-y\right) \Phi _{D}^{C}\left(
y,x\right) +W\left[ \Delta \right] ,  \label{104}
\end{equation}
\begin{eqnarray}
\exp \left\{ W\left[ \Delta \right] \right\} &=&1+\sum_{n=1}^{\infty }\Gamma
^{\left( n\right) }\left[ \Delta \right]  \nonumber \\
&=&\left\langle \text{T}\left[ \exp \left\{ -\int dx\int dy\overline{\psi }%
^{A}\left( x\right) \psi _{B}\left( y\right) \Delta _{A}^{B}\left(
x,y\right) \right\} \right] \right\rangle .\qquad  \label{105}
\end{eqnarray}
Calculations give 
\begin{equation}
\Gamma ^{\left( 1\right) }\left[ \Delta \right] =W^{\left( 1\right) }\left[
\Delta \right] =\int dx\int dy\,\Delta _{A}^{B}\left( x,y\right)
S_{B}^{A}\left( y-x\right) ,  \label{106}
\end{equation}
\begin{equation}
\Gamma ^{\left( 2\right) }\left[ \Delta \right] =\frac{1}{2}\left( W^{\left(
1\right) }\left[ \Delta \right] \right) ^{2}+W^{\left( 2\right) }\left[
\Delta \right] ,  \label{107}
\end{equation}
\begin{eqnarray}
W^{\left( 2\right) }\left[ \Delta \right] &=&-\frac{1}{2}\int dx_{1}\int
dy_{1}\int dx_{2}\int dy_{2}\Delta _{A_{1}}^{B_{1}}\left( x_{1},y_{1}\right)
S_{B_{1}}^{A_{2}}\left( y_{1}-x_{2}\right)  \nonumber \\
&&\Delta _{A_{2}}^{B_{2}}\left( x_{2},y_{2}\right) S_{B_{2}}^{A_{1}}\left(
y_{2}-x_{1}\right) ,  \label{108}
\end{eqnarray}
\begin{equation}
\Gamma ^{\left( 3\right) }\left[ \Delta \right] =\frac{1}{3!}\left(
W^{\left( 1\right) }\left[ \Delta \right] \right) ^{3}+W^{\left( 1\right)
}\left[ \Delta \right] W^{\left( 2\right) }\left[ \Delta \right] +W^{\left(
3\right) }\left[ \Delta \right] ,  \label{109}
\end{equation}
\begin{eqnarray}
W^{\left( 3\right) }\left[ \Delta \right] &=&\frac{1}{3}\int dx_{1}\int
dy_{1}...\,\int dx_{3}\int dy_{3}\Delta _{A_{1}}^{B_{1}}\left(
x_{1},y_{1}\right) S_{B_{1}}^{A_{2}}\left( y_{1}-x_{2}\right)  \nonumber \\
&&\Delta _{A_{2}}^{B_{2}}\left( x_{2},y_{2}\right) S_{B_{2}}^{A_{3}}\left(
y_{2}-x_{3}\right) \Delta _{A_{3}}^{B_{3}}\left( x_{3},y_{3}\right)
S_{B_{3}}^{A_{1}}\left( y_{3}-x_{1}\right) ,  \label{110} \\
&&..............\,\,.  \nonumber
\end{eqnarray}

\noindent \qquad The field equation is derived from the variational principle

\begin{equation}
\frac{\delta S_{\text{eff}}\left[ \Phi \right] }{\delta \Phi _{D}^{C}\left(
y,x\right) }=0
\end{equation}
and has the form

\begin{equation}
\Delta _{C}^{D}\left( y,x\right) =U_{CA}^{DB}\left( x-y\right) \frac{\delta
W\left[ \Delta \right] }{\delta \Delta _{A}^{B}\left( x,y\right) }.
\end{equation}
Using the expressions for $W^{\left( n\right) }\left[ \Delta \right] ,$ we
obtain the explicit equation 
\begin{eqnarray}
\Delta _{C}^{D}\left( y,x\right) &=&U_{CA}^{DB}\left( x-y\right) \left\{
S_{B}^{A}\left( y-x\right) \right.  \nonumber \\
&&-\int dx_{2}\int dy_{2}S_{B}^{A_{2}}\left( y-x_{2}\right) \Delta
_{A_{2}}^{B_{2}}\left( x_{2},y_{2}\right) S_{B_{2}}^{A}\left( y_{2}-x\right)
\nonumber \\
&&+\int dx_{2}\int dy_{2}\int dx_{3}\int dy_{3}S_{B}^{A_{2}}\left(
y-x_{2}\right) \Delta _{A_{2}}^{B_{2}}\left( x_{2},y_{2}\right)  \label{113}
\\
&&\hspace{1.3in}S_{B_{2}}^{A_{3}}\left( y_{2}-x_{3}\right) \Delta
_{A_{3}}^{B_{3}}\left( x_{3},y_{3}\right) S_{B_{3}}^{A}\left( y_{3}-x\right)
\nonumber \\
&&+\left. ...\hspace{3.5in}\QDATOP{\,}{\,}...\right\} .  \nonumber
\end{eqnarray}
Introduce the two-point Green function of quark field in the presence of the
quark-antiquark pairing 
\begin{eqnarray}
\mathbf{S}_{B}^{A}\left( y,x\right) &=&S_{B}^{A}\left( y-x\right) -\int
dx_{2}\int dy_{2}\,S_{B}^{A_{2}}\left( y-x_{2}\right) \Delta
_{A_{2}}^{B_{2}}\left( x_{2},y_{2}\right) S_{B_{2}}^{A}\left( y_{2}-x\right)
\nonumber \\
&&+\int dx_{2}\int dy_{2}\int dx_{3}\int dy_{3}\,S_{B}^{A_{2}}\left(
y-x_{2}\right) \Delta _{A_{2}}^{B_{2}}\left( x_{2},y_{2}\right)
S_{B_{2}}^{A_{3}}\left( y_{2}-x_{3}\right) \qquad  \nonumber \\
&&\qquad \Delta _{A_{3}}^{B_{3}}\left( x_{3},y_{3}\right)
S_{B_{3}}^{A}\left( y_{3}-x\right) +...  \label{114}
\end{eqnarray}
It satisfies the Schwinger-Dyson equation 
\begin{equation}
\mathbf{S}_{B}^{A}\left( y,x\right) =S_{B}^{A}\left( y-x\right) -\int dz\int
dw\,S_{B}^{C}\left( y-z\right) \Delta _{C}^{D}\left( z,w\right) \mathbf{S}%
_{D}^{A}\left( w,x\right) .  \label{115}
\end{equation}
Then the field equation (113) becomes 
\begin{equation}
\Delta _{C}^{D}\left( y,x\right) =U_{CA}^{DB}\left( x-y\right) \mathbf{S}%
_{B}^{A}\left( y,x\right) .  \label{116}
\end{equation}

Consider the solution of this field equation in the special class of
functions depending only on the difference of the coordinates 
\begin{equation}
\Delta _{A}^{B}\left( x,y\right) =\Delta _{A}^{B}\left( x-y\right) .
\label{117}
\end{equation}
In this case the function $\mathbf{S}_{B}^{A}\left( x,y\right) $ depends
also only on the difference of the coordinates 
\begin{equation}
\mathbf{S}_{B}^{A}\left( x,y\right) =\mathbf{S}_{B}^{A}\left( x-y\right) .
\label{118}
\end{equation}
Performing the Fourier transformation 
\begin{equation}
\Delta _{A}^{B}\left( \mathbf{x}-\mathbf{y},\tau -\sigma \right) =\frac{1}{%
\beta }\sum_{m}e^{i\varepsilon _{m}\left( \tau -\sigma \right) }\frac{1}{%
\left( 2\pi \right) ^{3}}\int d\mathbf{p}e^{i\mathbf{p}\left( \mathbf{x}-%
\mathbf{y}\right) }\widetilde{\Delta }_{A}^{B}\left( \mathbf{p},\varepsilon
_{m}\right) ,  \label{119}
\end{equation}
\[
\varepsilon _{m}=\left( 2m+1\right) \frac{\pi }{\beta }, 
\]
and introducing the matrices $\widetilde{\Delta }\left( \mathbf{p}%
,\varepsilon _{m}\right) $ with the elements $\widetilde{\Delta }%
_{A}^{B}\left( \mathbf{p},\varepsilon _{m}\right) $, we rewrite the integral
relation (114) and the integral equation (115) in the form of the algebraic
ones: 
\begin{eqnarray}
\widetilde{\mathbf{S}}\left( \mathbf{p},\varepsilon _{m}\right) &=&%
\widetilde{S}\left( \mathbf{p},\varepsilon _{m}\right) -\widetilde{S}\left( 
\mathbf{p},\varepsilon _{m}\right) \widetilde{\Delta }\left( \mathbf{p}%
,\varepsilon _{m}\right) \widetilde{S}\left( \mathbf{p},\varepsilon
_{m}\right)  \nonumber \\
&&+\widetilde{S}\left( \mathbf{p},\varepsilon _{m}\right) \widetilde{\Delta }%
\left( \mathbf{p},\varepsilon _{m}\right) \widetilde{S}\left( \mathbf{p}%
,\varepsilon _{m}\right) \widetilde{\Delta }\left( \mathbf{p},\varepsilon
_{m}\right) \widetilde{S}\left( \mathbf{p},\varepsilon _{m}\right) ...
\label{120}
\end{eqnarray}
and 
\begin{equation}
\widetilde{\mathbf{S}}\left( \mathbf{p},\varepsilon _{m}\right) =\widetilde{S%
}\left( \mathbf{p},\varepsilon _{m}\right) -\widetilde{S}\left( \mathbf{p}%
,\varepsilon _{m}\right) \widetilde{\Delta }\left( \mathbf{p},\varepsilon
_{m}\right) \widetilde{\mathbf{S}}\left( \mathbf{p},\varepsilon _{m}\right)
\label{121}
\end{equation}
or 
\begin{equation}
\frac{1}{\widetilde{\mathbf{S}}\left( \mathbf{p},\varepsilon _{m}\right) }=%
\frac{1}{\widetilde{S}\left( \mathbf{p},\varepsilon _{m}\right) }+\widetilde{%
\Delta }\left( \mathbf{p},\varepsilon _{m}\right) .  \label{122}
\end{equation}
The field equation (116) becomes 
\begin{equation}
\widetilde{\Delta }_{C}^{D}\left( \mathbf{p},\varepsilon _{m}\right) =\frac{1%
}{\beta }\sum_{n}\frac{1}{\left( 2\pi \right) ^{3}}\int d\mathbf{q}%
\widetilde{U}_{CA}^{DB}\left( \mathbf{p}-\mathbf{q},\varepsilon
_{m}-\varepsilon _{n}\right) \widetilde{\mathbf{S}}_{B}^{A}\left( \mathbf{q}%
,\varepsilon _{n}\right)  \label{123}
\end{equation}
where $\widetilde{U}_{CA}^{DB}\left( \mathbf{p}-\mathbf{q},\varepsilon
_{m}-\varepsilon _{n}\right) $ is the Fourier transform of $%
U_{CA}^{DB}\left( x-y\right) $, 
\begin{eqnarray}
U_{CA}^{DB}\left( \mathbf{x}-\mathbf{y},\tau -\sigma \right) &=&\frac{1}{%
\beta }\sum_{m}e^{i\omega _{m}\left( \sigma -\tau \right) }\frac{1}{\left(
2\pi \right) ^{3}}\int d\mathbf{p}e^{i\mathbf{p}\left( \mathbf{y}-\mathbf{x}%
\right) }\widetilde{U}_{CA}^{DB}\left( \mathbf{p},\omega _{m}\right) ,
\nonumber  \\
\omega _{m} &=&2m\frac{\pi }{\beta }.  \label{124}
\end{eqnarray}

Using the field equation (112), we obtain the value of the effective action
(104) 
\begin{eqnarray}
S_{eff}\left[ \Phi \right] &=&W\left[ \Delta \right] -\frac{1}{2}\int dx\int
dy\Delta _{A}^{B}\left( x,y\right) \frac{\delta \,W\left[ \Delta \right] }{%
\delta \,\Delta _{A}^{B}\left( x,y\right) }  \nonumber \\
&=&\left( 1-\frac{1}{2}\right) \int dx\int dy\Delta _{A}^{B}\left(
x,y\right) S_{B}^{A}\left( y-x\right)  \nonumber \\
&&+\left( \frac{1}{3}-\frac{1}{2}\right) \int dx_{1}\int dy_{1}...\,\int
dx_{3}\int dy_{3}\Delta _{A_{1}}^{B_{1}}\left( x_{1},y_{1}\right)
S_{B_{1}}^{A_{2}}\left( y_{1}-x_{2}\right)  \nonumber \\
&&\qquad \Delta _{A_{2}}^{B_{2}}\left( x_{2},y_{2}\right)
S_{B_{2}}^{A_{3}}\left( y_{2}-x_{3}\right) \Delta _{A_{3}}^{B_{3}}\left(
x_{3},y_{3}\right) S_{B_{3}}^{A_{1}}\left( y_{3}-x_{1}\right)  \nonumber \\
&&-\left( \frac{1}{4}-\frac{1}{2}\right) \int dx_{1}\int dy_{1}...\,\int
dx_{4}\int dy_{4}\Delta _{A_{1}}^{B_{1}}\left( x_{1},y_{1}\right)
S_{B_{1}}^{A_{2}}\left( y_{1}-x_{2}\right)  \nonumber \\
&&\qquad \Delta _{A_{2}}^{B_{2}}\left( x_{2},y_{2}\right) ...\,\Delta
_{A_{4}}^{B_{4}}\left( x_{4},y_{4}\right) S_{B_{4}}^{A_{1}}\left(
y_{4}-x_{1}\right)  \nonumber \\
&&+...\, \\
&=&\limfunc{Tr}\left\{ \left( 1-\frac{1}{2}\right) \int dx\int dy\Delta
\left( x,y\right) S\left( y-x\right) \right.  \nonumber \\
&&+\left( \frac{1}{3}-\frac{1}{2}\right) \int dx_{1}\int dy_{1}...\,\int
dx_{3}\int dy_{3}\,\Delta \left( x_{1},y_{1}\right) S\left(
y_{1}-x_{2}\right)  \nonumber \\
&&\,\,\qquad \,\Delta \left( x_{2},y_{2}\right) S\left( y_{2}-x_{3}\right)
\Delta \left( x_{3},y_{3}\right) S\left( y_{3}-x_{1}\right)  \nonumber \\
&&-\left( \frac{1}{4}-\frac{1}{2}\right) \int dx_{1}\int dy_{1}...\,\int
dx_{4}\int dy_{4}\Delta \left( x_{1},y_{1}\right) S\left( y_{1}-x_{2}\right)
\nonumber \\
&&\,\qquad \left. \QDATOP{\,}{\,}\Delta \left( x_{2},y_{2}\right)
...\,\Delta \left( x_{4},y_{4}\right) S\left( y_{4}-x_{1}\right) +...\hspace{%
1.4in}\right\} ,  \nonumber
\end{eqnarray}
where $\Delta \left( x,y\right) $ is the matrix with the elements $\Delta
_{A}^{B}\left( x,y\right) $. For the fields in the special class (117) the
effective action is expressed in terms of the free energy density $F\left[ 
\mathbf{x};\Delta \right] $%
\begin{equation}
S_{eff}\left[ \Phi \right] =-\beta \int d\mathbf{x}F\left[ \mathbf{x};\Delta
\right] .  \label{126}
\end{equation}
It follows that\nolinebreak \nolinebreak \- 
\begin{eqnarray}
F\left[ \mathbf{x};\Delta \right] &=&F\left[ \Delta \right] =  \nonumber \\
&=&\left( 1-\frac{1}{2}\right) \int dy\Delta _{A}^{B}\left( x-y\right)
S_{B}^{A}\left( y-x\right)  \nonumber \\
&&+\left( \frac{1}{3}-\frac{1}{2}\right) \int dy\int dx_{2}\int dy_{2}\,\int
dx_{3}\int dy_{3}\Delta _{A_{1}}^{\,B_{1}}\left( x-y_{1}\right)
S_{B_{1}}^{A_{2}}\left( y_{1}-x_{2}\right)  \nonumber \\
&&\qquad \Delta _{A_{2}}^{B_{2}}\left( x_{2}-y_{2}\right)
S_{B_{2}}^{A_{3}}\left( y_{2}-x_{3}\right) \Delta _{A_{3}}^{B_{3}}\left(
x_{3}-y_{3}\right) S_{B_{3}}^{A_{1}}\left( y_{3}-x\right)  \nonumber \\
&&-\left( \frac{1}{4}-\frac{1}{2}\right) \,\int dy\int dx_{2}\int
dy_{2}...\,\int dx_{4}\int dy_{4}\Delta _{A_{1}}^{B_{1}}\left(
x-y_{1}\right) S_{B_{1}}^{A_{2}}\left( y_{1}-x_{2}\right)  \nonumber \\
&&\qquad \Delta _{A_{2}}^{B_{2}}\left( x_{2}-y_{2}\right)
...\,S_{B_{3}}^{A_{4}}\left( y_{3}-x_{4}\right) \Delta
_{A_{4}}^{B_{4}}\left( x_{4}-y_{4}\right) S_{B_{4}}^{A_{1}}\left(
y_{4}-x\right)  \nonumber \\
&&+...\,  \nonumber \\
&=&\limfunc{Tr}\left\{ \left( 1-\frac{1}{2}\right) \int dy\Delta \left(
x-y\right) S\left( y-x\right) \right.  \nonumber \\
&&+\left( \frac{1}{3}-\frac{1}{2}\right) \int dy\int dx_{2}\int dy_{2}\,\int
dx_{3}\int dy_{3}\Delta \left( x-y_{1}\right) S\left( y_{1}-x_{2}\right) 
\nonumber \\
&&\qquad \Delta \left( x_{2}-y_{2}\right) S\left( y_{2}-x_{3}\right) \Delta
\left( x_{3}-y_{3}\right) S\left( y_{3}-x\right)  \nonumber \\
&&-\left( \frac{1}{4}-\frac{1}{2}\right) \,\int dy\int dx_{2}\int
dy_{2}...\,\int dx_{4}\int dy_{4}\Delta \left( x-y_{1}\right) S\left(
y_{1}-x_{2}\right)  \nonumber \\
&&\qquad \left. \QDATOP{\,}{\,}\Delta \left( x_{2}-y_{2}\right) ...\,S\left(
y_{3}-x_{4}\right) \Delta \left( x_{4}-y_{4}\right) S\left( y_{4}-x\right)
+...\right\}  \label{127} \\
&=&\frac{1}{\beta }\sum_{m}\frac{1}{\left( 2\pi \right) ^{3}}\int d\mathbf{p}%
\limfunc{Tr}\left\{ \left( 1-\frac{1}{2}\right) \widetilde{\Delta }\left( 
\mathbf{p},\varepsilon _{m}\right) \widetilde{S}\left( \mathbf{p}%
,\varepsilon _{m}\right) \right.  \nonumber \\
&&+\left( \frac{1}{3}-\frac{1}{2}\right) \left[ \widetilde{\Delta }\left( 
\mathbf{p},\varepsilon _{m}\right) \widetilde{S}\left( \mathbf{p}%
,\varepsilon _{m}\right) \right] ^{3}  \nonumber \\
&&\left. -\left( \frac{1}{4}-\frac{1}{2}\right) \left[ \widetilde{\Delta }%
\left( \mathbf{p},\varepsilon _{m}\right) \widetilde{S}\left( \mathbf{p}%
,\varepsilon _{m}\right) \right] ^{4}+...\,\hspace{1.5in}\right\} . 
\nonumber
\end{eqnarray}
Summing up the infinite series, we obtain

\begin{equation}
F\left[ \Delta \right] =\frac{1}{\beta }\sum_{m}\frac{1}{\left( 2\pi \right)
^{3}}\int d\mathbf{p}\limfunc{Tr}\left\{ \widetilde{\Delta }\left( \mathbf{p}%
,\varepsilon _{m}\right) \left[ \int_{0}^{1}\widetilde{\mathbf{S}}^{\xi
}\left( \mathbf{p},\varepsilon _{m}\right) d\xi -\frac{1}{2}\widetilde{%
\mathbf{S}}\left( \mathbf{p},\varepsilon _{m}\right) \right] \right\} ,
\label{128}
\end{equation}
where $\widetilde{\mathbf{S}}^{\xi }\left( \mathbf{p},\varepsilon
_{m}\right) $ is determined by the equation of the form (122) with the
replacement of $\widetilde{\Delta }\left( \mathbf{p},\varepsilon _{m}\right) 
$ by $\xi \widetilde{\Delta }\left( \mathbf{p},\varepsilon _{m}\right) $

\begin{equation}
\frac{1}{\widetilde{\mathbf{S}}^{\xi }\left( \mathbf{p},\varepsilon
_{m}\right) }=\frac{1}{\widetilde{\mathbf{S}}\left( \mathbf{p},\varepsilon
_{m}\right) }+\xi \widetilde{\Delta }\left( \mathbf{p},\varepsilon
_{m}\right) .  \label{129}
\end{equation}
Introduce $4$-vector $p_{\mu }$ with $p_{4}=\varepsilon _{m}$ and denote $%
\widetilde{\Delta }_{A}^{B}\left( \mathbf{p},\varepsilon _{m}\right) $by $%
\widetilde{\Delta }_{A}^{B}\left( p\right) .$ These order parameters have
following most general form

\begin{eqnarray}
\widetilde{\Delta }_{A}^{B}\left( p\right) &=&\delta _{\alpha }^{\beta }%
\widetilde{\Delta }_{\left( ai\right) }^{S\left( bj\right) }\left( p\right)
+\left( \gamma _{5}\right) _{\alpha }^{\beta }\widetilde{\Delta }_{\left(
ai\right) }^{P\left( bj\right) }\left( p\right) +\left( \gamma _{\mu
}\right) _{\alpha }^{\beta }\widetilde{\Delta }_{\mu \left( ai\right)
}^{V\left( bj\right) }\left( p\right)  \nonumber \\
&&+\left( \gamma _{\mu }\gamma _{5}\right) _{\alpha }^{\beta }\widetilde{%
\Delta }_{\mu \left( ai\right) }^{A\left( bj\right) }\left( p\right) +\left(
\sigma _{\mu \nu }\right) _{\alpha }^{\beta }\widetilde{\Delta }_{\mu \nu
\left( ai\right) }^{t\left( bj\right) }.\left( p\right)  \label{130}
\end{eqnarray}
The existence of the non-vanishing order parameters with definite
transformation properties would mean the spontaneous breaking of the
corresponding symmetries.

\section{Formation of triquarks}

It is straightforward to generalize the method presented in preceding
Sections for applying to the problem of the formation of triquarks - the
bound states of three quarks. We note that the fundamental interaction
mechanisms in QCD (instanton induced, gluon exchange, etc...) always lead to
some effective (non-local, in general) six-fermion interaction between quark
fields with the effective interaction action of the general form

\begin{eqnarray*}
S_{\text{int}} &=&\frac{1}{6}\int dx\int dy\int dz\int du\int dv\int dw%
\overline{\psi }^{F}\left( w\right) \overline{\psi }^{E}\left( v\right) 
\overline{\psi }^{D}\left( u\right) \\
&&\,\,\,\,\,\,\,\,\,\,\,\,\,\,\,\,\,\,\,\,\,\,\,\,\,\,\,\,\,\,\,\,%
\,V_{DEF}^{CBA}\left( u,v,w;z,y,x\right) \psi _{A}\left( x\right) \psi
_{B}\left( y\right) \psi _{C}\left( z\right) ,\,\,
\end{eqnarray*}
\begin{equation}
\qquad \,V_{DEF}^{CBA}\left( u,v,w;z,y,x\right) =-V_{DEF}^{CAB}\left(
u,v,w;z,y,x\right) =-V_{EDF}^{CBA}\left( u,v,w;z,y,x\right) =...  \label{131}
\end{equation}
The form-factors $\,\,V_{DEF}^{CBA}\left( u,v,w;z,y,x\right) $ depend only
on the differences of the space-time coordinates. The partition function of
the many-quark system with this effective six-quark interaction equals

\begin{eqnarray}
Z &=&\int \left[ D\psi \right] \left[ D\overline{\psi }\right] \exp \left\{
-\int dx\overline{\psi }^{A}\left( x\right) L_{A}^{B}\psi _{B}\left(
x\right) \right\}  \nonumber \\
&&\exp \left\{ \frac{1}{6}\int dx\int dy\int dz\int du\int dv\int dw%
\overline{\psi }^{F}\left( w\right) \overline{\psi }^{E}\left( v\right) 
\overline{\psi }^{D}\left( u\right) \right.  \nonumber \\
&&\,\,\,\,\,\,\,\,\,\,\,\,\,\left. \,V_{DEF}^{CBA}\left( u,v,w;z,y,x\right)
\psi _{A}\left( x\right) \psi _{B}\left( y\right) \psi _{C}\left( z\right)
\right\} .  \label{132}
\end{eqnarray}

In order to describe the triquarks we introduce some tri-spinor tri-local
field $\Phi _{ABC}\left( x,y,z\right) $ as well as its conjugate $\overline{%
\Phi }^{CBA}\left( z,y,x\right) $ and set

\begin{eqnarray}
Z_{0}^{\Phi } &=&\int \left[ D\Phi \right] \left[ D\overline{\Phi }\right]
\exp \left\{ \frac{1}{6}\int dx\int dy\int dz\int du\int dv\int dw\right. 
\nonumber \\
&&\left. \overline{\Phi }^{FED}\left( w,v,u\right) V_{DEF}^{CBA}\left(
u,v,w;z,y,x\right) \Phi _{ABC}\left( x,y,z\right) \right\} .  \label{133}
\end{eqnarray}
By shifting the functional integration variables

\begin{eqnarray*}
\Phi _{ABC}\left( x,y,z\right) &\rightarrow &\Phi _{ABC}\left( x,y,z\right) +%
\frac{1}{\sqrt{2}}\psi _{A}\left( x\right) \psi _{B}\left( y\right) \psi
_{C}\left( z\right) , \\
\overline{\Phi }^{FED}\left( w,v,u\right) &\rightarrow &\overline{\Phi }%
^{FED}\left( w,v,u\right) +\frac{1}{\sqrt{2}}\overline{\psi }^{F}\left(
w\right) \overline{\psi }^{E}\left( v\right) \overline{\psi }^{D}\left(
u\right) ,
\end{eqnarray*}
we establish the Hubbard-Stratonovich transformation

\begin{eqnarray}
&&\exp \left\{ \frac{1}{6}\int dx\int dy\int dz\int du\int dv\int dw%
\overline{\psi }^{F}\left( w\right) \overline{\psi }^{E}\left( v\right) 
\overline{\psi }^{D}\left( u\right) \right.  \nonumber \\
&&\;\;\;\;\;\;\;\;\;\;\;\;\;\;\;\;\;\left. V_{DEF}^{CBA}\left(
u,v,w;z,y,x\right) \psi _{A}\left( x\right) \psi _{B}\left( y\right) \psi
_{C}\left( z\right) \right\}  \nonumber \\
&=&\frac{1}{Z_{0}^{\Phi }}\int \left[ D\Phi \right] \left[ D\overline{\Phi }%
\right] \exp \left\{ -\frac{1}{3}\int dx\int dy\int dz\int du\int dv\int
dw\right.  \nonumber \\
&&\,\,\,\,\,\,\,\,\,\,\,\,\,\,\,\,\left. \overline{\Phi }^{FED}\left(
w,v,u\right) V_{DEF}^{CBA}\left( u,v,w;z,y,x\right) \Phi _{ABC}\left(
x,y,z\right) \right\}  \nonumber \\
&&\exp \left\{ -\frac{1}{\sqrt{18}}\int dx\int dy\int dz\left[ \overline{%
\Delta }^{CBA}\left( z,y,x\right) \psi _{A}\left( x\right) \psi _{B}\left(
y\right) \psi _{C}\left( x\right) \right. \right.  \nonumber \\
&&\,\,\,\,\,\,\,\,\,\,\,\,\,\,\,\,\,\,\,\,\left. \left. +\overline{\psi }%
^{C}\left( z\right) \overline{\psi }^{B}\left( y\right) \overline{\psi }%
^{A}\left( x\right) \Delta _{ABC}\left( x,y,z\right) \right] \right\} ,
\label{134}
\end{eqnarray}
where

\begin{eqnarray}
\overline{\Delta }^{CBA}\left( z,y,x\right) &=&\int du\int dv\int dw%
\overline{\psi }^{F}\left( w\right) \overline{\psi }^{E}\left( v\right) 
\overline{\psi }^{D}\left( u\right) V_{DEF}^{CBA}\left( u,v,w;z,y,x\right) ,
\label{135} \\
\Delta _{ABC}\left( x,y,z\right) &=&\int du\int dv\int dwV_{DEF}^{CBA}\left(
u,v,w;z,y,x\right) \psi _{D}\left( u\right) \psi _{E}\left( v\right) \psi
_{F}\left( w\right) ,\hskip 0.8cm  \nonumber
\end{eqnarray}
and transform, after lengthy calculations, the partition function (132) into
the form (23) of a functional integral over the tri-local fields $\Phi
_{ABC}\left( x,y,z\right) $ and $\overline{\Phi }^{CBA}\left( z,y,x\right) $
with the effective action

\begin{eqnarray}
S_{\text{eff}}\left[ \Phi ,\overline{\Phi }\right] &=&-\frac{1}{3}\int
dx\int dy\int dz\overline{\Phi }^{FDE}\left( w,v,u\right)
V_{DEF}^{CBA}\left( u,v,w;z,y,x\right) \Phi _{ABC}\left( x,y,z\right) 
\nonumber \\
&&+W\left[ \Delta ,\overline{\Delta }\right]  \label{136}
\end{eqnarray}
and the functional $W\left[ \Delta ,\overline{\Delta }\right] $ of the form

\begin{equation}
W\left[ \Delta ,\overline{\Delta }\right] =\sum_{n=1}^{\infty }W^{\left(
2n\right) }\left[ \Delta ,\overline{\Delta }\right] ,  \label{137}
\end{equation}
where $W^{\left( 2n\right) }\left[ \Delta ,\overline{\Delta }\right] $ is a
functional of the $n-$th order with respect to each type of tri-local fields 
$\Delta _{ABC}\left( x,y,z\right) $and $\overline{\Delta }^{CBA}\left(
z,y,x\right) $, for example

\begin{eqnarray}
W^{\left( 2\right) }\left[ \Delta ,\overline{\Delta }\right] &=&\frac{1}{3}%
\int dx\int dy\int dz\int du\int dv\int dw  \label{138}  \\
&&\overline{\Delta }^{FED}\left( w,v,u\right) S_{F}^{C}\left( w-z\right)
S_{E}^{B}\left( v-y\right) S_{D}^{A}\left( u-x\right) \Delta _{ABC}\left(
x,y,z\right) ,  \nonumber
\end{eqnarray}

\begin{eqnarray}
W^{\left( 4\right) }\left[ \Delta ,\overline{\Delta }\right] &=&-\frac{1}{2}%
\int dx_{1}\int dy_{1}\int dz_{1}...\int du_{2}\int dv_{2}\int dw_{2}%
\overline{\Delta }^{F_{1}E_{1}D_{1}}\left( w_{1},v_{1},u_{1}\right) 
\nonumber \\
&&\overline{\Delta }^{F_{2}E_{2}D_{2}}\left( w_{2},v_{2},u_{2}\right)
S_{F_{2}}^{C_{1}}\left( w_{2}-z_{1}\right) S_{E_{1}}^{B_{1}}\left(
v_{1}-y_{1}\right) S_{D_{2}}^{A_{1}}\left( u_{1}-x_{1}\right)
S_{F_{1}}^{C_{2}}\left( w_{1}-z_{2}\right) \qquad  \nonumber \\
&&S_{E_{2}}^{B_{2}}\left( v_{2}-y_{2}\right) S_{D_{2}}^{A_{2}}\left(
u_{2}-x_{2}\right) \Delta _{A_{2}B_{2}C_{2}}\left( x_{2},y_{2},z_{2}\right)
\Delta _{A_{1}B_{1}C_{1}}\left( x_{1},y_{1},z_{1}\right)  \label{139} \\
&&..............  \nonumber
\end{eqnarray}
From the variational principle we derive the field equation

\begin{equation}
\frac{1}{3}\Delta _{ABC}\left( x,y,z\right) =\int du\int dv\int
dwV_{ABC}^{FED}\left( u,v,w;z,y,x\right) \sum_{n=1}^{\infty }\frac{\delta
W^{\left( n\right) }\left[ \Delta ,\overline{\Delta }\right] }{\delta 
\overline{\Delta }^{FED}\left( w,v,u\right) }.  \label{140}
\end{equation}
Using explicit expression of $W^{\left( 2n\right) }\left[ \Delta ,\overline{%
\Delta }\right] $, we have shown that up to the $6$th order of the
pertubation theory there exists following system of integral equations

\begin{eqnarray}
\Delta _{ABC}\left( x,y,z\right) &=&\int du\int dv\int dw\int dx^{\prime
}\int dy^{\prime }\int dz^{\prime }V_{ABC}^{FED}\left( u,v,w;z,y,x\right) 
\nonumber \\
&&G_{F}^{C^{\prime }}\left( w,z^{\prime }\right) G_{E}^{B^{\prime }}\left(
v,y^{\prime }\right) G_{D}^{A^{\prime }}\left( u,x^{\prime }\right) \Delta
_{A^{\prime }B^{\prime }C^{\prime }}\left( x^{\prime },y^{\prime },z^{\prime
}\right) ,  \label{141}
\end{eqnarray}

\begin{equation}
G_{A}^{D}\left( x,u\right) =S_{A}^{D}\left( x-u\right) +\int dy\int
dzS_{A}^{B}\left( x-y\right) \Sigma _{B}^{C}\left( y,z\right)
G_{C}^{D}\left( z,u\right) ,  \label{142}
\end{equation}

\begin{equation}
\Sigma _{C}^{F}\left( z,w\right) =\int dx\int dy\int du\int dv\overline{%
\Delta }^{FED}\left( w,v,u\right) G_{D}^{A}\left( u,x\right) G_{E}^{B}\left(
v,y\right) \Delta _{ABC}\left( x,y,z\right) .  \label{143}
\end{equation}
It is easy to verify that $G_{A}^{D}\left( x,u\right) $ is the two-point
Green function of the quark field in the presence of its interaction with
the ''external'' tri-local fields $\Delta _{ABC}\left( x,y,z\right) $ and $%
\overline{\Delta }^{FED}\left( w,v,u\right) $. It is determined by the
Schwinger-Dyson equation represented by the Feynman diagram in Fig. 1. with
the self-energy part (143) reprsented by the Feynman diagram in Fig. 2.\par
The solutions of the system of integral equations (141)-(143) in the class
of the fields depending only on the differences of the coordinates can be
considered as the anticommuting order parameters of the ground state of the
many-quark system with the binding of the quarks into the triquarks. This
means that in the QCD dense quark matter there might exist a phase
transition with the anticommuting order parameters. Note that there is no
condensation of the triquarks, because these composite particles are
fermions.

\vskip 0.4cm%
{\Large \bf Acknowledgements.}%

\vskip 0.3cm%
The author would like to express his sincere appreciation to the National
Natural Sciences Council of Vietnam for the support to this work.%
\vskip 0.4cm%

{\Large \bf References.}%

\begin{itemize}
\item[{\lbrack 1].}]  B. C. Barrois, \textit{Nucl. Phys.} \textbf{B129}
(1977) 390.

\item[{\lbrack 2].}]  C. Frautschi, \textit{Asymptotic Freedom and Color
Superconductivity in Dense Quark Matter, in: Proceedings of the Workshop on
Hadronic Matter at Extreme Energy Density}, Ed., N. Cabibbo, Erice, Italy
(1978).

\item[{\lbrack 3].}]  D. Bailin and A. Love, \textit{Nucl. Phys.} \textbf{%
B190} (1981) 175; \textbf{B190} (1981) 751; \textbf{B205} (1982) 119.

\item[{\lbrack 4].}]  J. F. Donoghue and K. S. Sateesh, \textit{Phys. Rev.} 
\textbf{D38} (1988) 360.

\item[{\lbrack 5].}]  M. Iwasaki and T. Iwado, \textit{Phys. Lett.} \textbf{%
B350} (1995) 163.

\item[{\lbrack 6].}]  M. Alford, K. Rajagopal and F. Wilczek , \textit{Phys.
Lett. }\textbf{B422} (1998) 247; \textit{Nucl. Phys},. \textbf{B537} (1999)
443.

\item[{\lbrack 7].}]  T. Sch\"{a}fer and F. Wilczek, \textit{Phys. Rev. Lett.%
} \textbf{82} (1999) 3956; \textit{Phys. Lett.} \textbf{B450} (1999) 325.

\item[{\lbrack 8].}]  R. Rapp, T. Sch\"{a}fer, E. Shuryak and M. Velkovsky, 
\textit{Phys. Rev. Lett.},\textbf{\ 81} (1998) 53.

\item[{\lbrack 9].}]  N.Evans, S. Hsu and M. Schwetz, \textit{Nucl. Phys.} 
\textbf{B551} (1999) 275; \textit{Phys. Lett.} \textbf{B449} (1999) 281.

\item[{\lbrack 10].}]  D. T. Son, \textit{Phys. Rev.} \textbf{D59 }(1998)
094019.

\item[{\lbrack 11].}]  G. W. Carter and D. Diakonov, \textit{Phys. Rev}. 
\textbf{D60 }(1999) 01004.

\item[{\lbrack 12].}]  R. Rapp, T. Sch\"{a}fer, E. Shuryak and M. Velkovsky,
preprint IASSNS-HEP-99/40, Princeton, 1999, hep-ph/9904353.

\item[{\lbrack 13].}]  A. Chodos, H. Minakata and F. Cooper, \textit{Phys.
Lett}. \textbf{B449} (1999) 260.

\item[{\lbrack 14].}]  R. D. Pisarski and D. H. Rischke, \textit{Phys. Rev}. 
\textbf{D60} (1999) 094013.

\item[{\lbrack 15].}]  T Sch\"{a}fer and F. Wilczek, \textit{Phys. Rev}. 
\textbf{D60} (1999) 074014, 114033.

\item[{\lbrack 16].}]  V. A. Miransky, I. A. Shovkovy and L. C.
Wijewardhana, \textit{\ Phys. Lett. }\textbf{B468} (1999) 270.

\item[{\lbrack 17].}]  R. Casalbuoni and R. Gatto, \textit{Phys. Lett}. 
\textbf{B464} (1999) 111.

\item[{\lbrack 18].}]  D. K. Hong, M. Rho and I. Zahed, \textit{Phys. Lett}. 
\textbf{B468} (1999) 261.

\item[{\lbrack 19].}]  M. Rho, E. Shuryak, A. Wirzba and I. Zahed, \textit{%
hep-/0001104}.

\item[{\lbrack 20].}]  C. Manuel and M. H. G. Tytgat, \textit{hep-ph/0001095.%
}

\item[{\lbrack 21].}]  W. Brown, J. T. Liu and H. C. Ren, \textit{%
hep-ph/0003199.}

\item[{\lbrack 22].}]  J. Berges and K. Rajagopal, \textit{Nucl. Phys.} 
\textbf{B358} (1999) 215.

\item[{\lbrack 23].}]  M. Harada and A. Shibata, \textit{Phys. Rev.} \textbf{%
D59} (1998) 014010.

\item[{\lbrack 24].}]  R. D. Pisarski and D. H. Rischke, \textit{Phys. Rev.
Lett}. \textbf{83} (1999) 37.

\item[{\lbrack 25].}]  Nguyen Van Hieu, \textit{Basics of Fuctional Integral
Technique in Quantum Field Theory of Many-Body Systems, VNUH Pub}., Hanoi,
1999.

\item[{\lbrack 26].}]  Nguyen Van Hieu and Le Trong Tuong, \textit{Commun.
Phys.} \textbf{8} (1998) 129.

\item[{\lbrack 27].}]  Nguyen Van Hieu, Nguyen Hung Son, Ngo Van Thanh and
Hoang Ba Thang, \textit{Adv . Nat. Sci}. \textbf{1} (2000) 61, \textit{%
hep-ph/0001251.}

\item[{\lbrack 28].}]  Nguyen van Hieu, Hoang Ngoc Long, Nguyen Hung Son and
Nguyen Ai Viet, \textit{Proc. 24th Conference in Theoretical Physics, Samson
19-21 August 1999, p.1}, \textit{hep-ph/0001234.}

\item[{\lbrack 29].}]  Nguyen Van Hieu., \textit{Functional Integral
Techniques in Condensed Matter Physics}, in the book \textit{Computational
Approaches for Novel Condensed Matter Systems}, Ed. Mukunda Das, Plenum
Press, New York, 1994, p. 194-234.

\item[{\lbrack 30].}]  Nguyen Van Hieu., Aus. J. Phys. \textbf{50}
(1997)1035.
\end{itemize}

\end{document}